\documentclass[journal]{IEEEtran}

\usepackage{cite,graphicx,epstopdf,float,url,amssymb,amsthm,threeparttable,multirow,algorithmic,nameref,paralist}
\usepackage{amsfonts}
\usepackage[labelformat=simple]{subcaption}

\usepackage[ruled,vlined]{algorithm2e}
\usepackage[usenames]{color}
\usepackage[normalem]{ulem}
\usepackage[cmex10]{amsmath}

\usepackage{diagbox,array}
\newcolumntype{C}[1]{>{\centering\let\newline\\\arraybackslash\hspace{0pt}}m{#1}}

\usepackage{etoolbox}

\newcounter{numrel}
\renewcommand{\thenumrel}{\alph{numrel}}

\makeatletter
\newcommand{\numrel}[2]{
  \refstepcounter{numrel}
  \ltx@label{#2}
  \overset{(\thenumrel)}{#1}
}
\makeatother
\AfterEndEnvironment{IEEEproof}{\setcounter{numrel}{0}}

\theoremstyle{plain}
\newtheorem{lemma}{Lemma}
\newtheorem{cor}{Corollary}
\newtheorem{theorem}{Theorem}

\theoremstyle{definition}

\theoremstyle{remark}
\newtheorem{remark}{Remark}

\usepackage{url}
\usepackage[dvipsnames]{xcolor}
\usepackage[colorinlistoftodos,color=orange!70,textsize=scriptsize,shadow]{todonotes}
\usepackage{setspace}

\usepackage[normalem]{ulem}
\newcommand\redout{\bgroup\markoverwith{\textcolor{red}{\rule[.5ex]{2pt}{0.4pt}}}\ULon}

\newcommand{\mb}{\mathbf}
\newcommand{\mc}{\mathcal}

\newcommand{\mbb}{\mathbb}

\newcommand{\A}{\mb{A}}
\newcommand{\B}{\mb{B}}
\newcommand{\D}{\mb{D}}
\newcommand{\X}{\mb{X}}
\newcommand{\Y}{\mb{Y}}
\newcommand{\N}{\mb{N}}
\newcommand{\T}{\mb{T}}
\newcommand{\I}{\mb{I}}
\newcommand{\U}{\mb{U}}
\newcommand{\G}{\mb{G}}
\newcommand{\Delt}{\mb{\Delta}}
\newcommand{\Sig}{\mb{\Sigma}}

\newcommand{\x}{\mb{x}}
\newcommand{\y}{\mb{y}}

\newcommand{\bi}{\mb{i}}
\newcommand{\ba}{\mb{a}}
\newcommand{\bb}{\mb{b}}
\newcommand{\bd}{\mb{d}}
\newcommand{\bu}{\mb{u}}
\newcommand{\bv}{\mb{v}}
\newcommand{\bg}{\mb{g}}

\newcommand{\bbR}{\mbb{R}}
\newcommand{\bbE}{\mbb{E}}
\newcommand{\bbP}{\mbb{P}}

\newcommand{\cS}{\mc{S}}
\newcommand{\cD}{\mc{D}}

\newcommand{\cX}{\mc{X}}
\newcommand{\cE}{\mc{E}}
\newcommand{\cL}{\mc{L}}

\newcommand{\ul}{\underline}
\newcommand{\uY}{\ul{\Y}}
\newcommand{\uX}{\ul{\X}}
\newcommand{\uN}{\ul{\N}}

\newcommand{\wh}{\widehat}
\newcommand{\wt}{\widetilde}
\newcommand{\eps}{\varepsilon}

\newcommand{\al}{\alpha}

\newcommand{\et}{\eta}
\newcommand{\no}{\nu}

\newcommand{\vect}{\mathop{\mathrm{vec}}\nolimits}
\newcommand{\SNR}{\mathop{\mathrm{SNR}}\nolimits}
\newcommand{\supp}{\mathop{\mathrm{supp}}\nolimits}
\newcommand{\tr}{\mathop{\mathrm{Tr}}\nolimits}
\newcommand{\rnk}{\mathop{\mathrm{rank}}\nolimits}
\newcommand{\RIP}{\mathsf{RIP}}

\newcommand{\norm}[1]{ \left\| #1 \right\| }

\newcommand{\ip}[2]{ \left\langle #1, #2 \right\rangle }
\newcommand{\lr}[1]{ \left\{ #1 \right\} }

\newcommand{\lrp}[1]{ \left( #1 \right) }
\newcommand{\lra}[1]{ \left| #1 \right| }
\newcommand{\flr}[1]{ \lfloor #1 \rfloor }

\DeclareMathOperator*{\bigAst}{\raisebox{-0.6ex}{\scalebox{2.5}{$\ast$}}}

\IEEEoverridecommandlockouts

\begin{document}
\title{Minimax Lower Bounds on Dictionary Learning for Tensor Data}
\author{%
Zahra~Shakeri
,
Waheed~U.~Bajwa
, and
Anand~D.~Sarwate
\thanks{Manuscript received August 29, 2016; revised October 3, 2017 and January 12, 2018; accepted January 14, 2018. This work is supported in part by the National Science Foundation under awards CCF-1525276 and CCF-1453073, and by the Army Research Office under awards W911NF-14-1-0295 and W911NF-17-1-0546. Some of the results reported here were presented at the 2016 IEEE International Symposium on Information Theory (ISIT)~\cite{shakeri2016minimax} and at the 2017 IEEE International Conference on Acoustics, Speech, and Signal Processing (ICASSP)~\cite{shakeri2017sample}.
}
\thanks{The authors are with the Department of Electrical and Computer Engineering, Rutgers, The State University of New Jersey, 94 Brett Road, Piscataway, NJ 08854, USA.  (Emails: {\tt zahra.shakeri@rutgers.edu}, {\tt
waheed.bajwa@rutgers.edu}, and  {\tt anand.sarwate@rutgers.edu})}
\thanks{Copyright (c) 2018 IEEE. Personal use of this material is permitted. However, permission to use this material for any other purposes must be obtained from the IEEE by sending a request to \texttt{pubs-permissions@ieee.org}.}
}

\maketitle

\begin{abstract}
This paper provides fundamental limits on the sample complexity of estimating dictionaries for tensor data. The specific focus of this work is on $K$th-order tensor data and the case where the underlying dictionary can be expressed in terms of $K$ smaller dictionaries. It is assumed the data are generated by linear combinations of these structured dictionary atoms and observed through white Gaussian noise. This work first provides a general lower bound on the minimax risk of dictionary learning for such tensor data and then adapts the proof techniques for specialized results in the case of sparse and sparse-Gaussian linear combinations. The results suggest the sample complexity of dictionary learning for tensor data can be significantly lower than that for unstructured data: for unstructured data it scales linearly with the product of the dictionary dimensions, whereas for tensor-structured data the bound scales linearly with the sum of the product of the dimensions of the (smaller) component dictionaries. A partial converse is provided for the case of 2nd-order tensor data to show that the bounds in this paper can be tight. This involves developing an algorithm for learning highly-structured dictionaries from noisy tensor data. Finally, numerical experiments highlight the advantages associated with explicitly accounting for tensor data structure during dictionary learning.
\end{abstract}

\begin{IEEEkeywords}
Dictionary learning, Kronecker-structured dictionary, minimax bounds, sparse representations, tensor data.
\end{IEEEkeywords}

\section{Introduction}\label{sec:Introduction}

Dictionary learning is a technique for finding sparse representations of signals or data and has applications in various tasks such as image denoising and inpainting~\cite{aharon2006img}, audio processing~\cite{grosse2012shift}, and classification~\cite{raina2007self,mairal2012task}. Given input training signals $\lr{ \y_n \in \bbR^m }_{n=1}^N$, the goal in dictionary learning is to construct an overcomplete basis, $\D \in \bbR^{m \times p}$, such that each signal in $\Y=\big[ \y_1, \dots , \y_N \big]$ can be described by a small number of atoms (columns) of $\D$~\cite{kreutz2003dictionary}.
This problem can be posed as the following optimization program:
\begin{align}
\min_{\D,\X} \|\Y - \D \X\|_F \quad \text{subject to} \ \forall n, \|\x_n\|_0 \leq s,
\end{align}
where $\x_n$ is the coefficient vector associated with $\y_n$, $\| \cdot \|_0$ counts the number of nonzero entries and $s$ is the maximum number of nonzero elements of $\x_n$. Although existing literature has mostly focused on dictionary learning for one-dimensional data~\cite{aharon2006img,grosse2012shift,raina2007self,mairal2012task, kreutz2003dictionary}, many real-world signals are multidimensional and have a tensor structure: examples include images, videos, and signals produced via magnetic resonance or computed tomography systems. In traditional dictionary learning literature, multidimensional data are converted into one-dimensional data by vectorizing the signals. Such approaches can result in poor sparse representations because they neglect the multidimensional structure of the data~\cite{zhang2015denoising}. This suggests that it might be useful to keep the original tensor structure of multidimensional data for efficient dictionary learning and reliable subsequent processing.

There have been several algorithms proposed in the literature that can be used to learn structured dictionaries for multidimensional data~\cite{hawe2013separable,zubair2013tensor,
roemer2014tensor,dantas2017learning,ghassemi2017stark,peng2014decomposable,soltani2015tensor,duan2012k,zhang2015denoising}. In~\cite{hawe2013separable}, a Riemannian conjugate gradient method combined with a nonmonotone line search is used to learn structured dictionaries. Other structured dictionary learning works rely on various tensor decomposition methods such as the Tucker decomposition~\cite{tucker1963implications, zubair2013tensor,peng2014decomposable,dantas2017learning,ghassemi2017stark}, the CANDECOMP/PARAFAC (CP) decomposition~\cite{harshman1970foundations,duan2012k}, the HOSVD decomposition~\cite{de2000multilinear,roemer2014tensor}, the t-product tensor
factorization~\cite{soltani2015tensor}, and the tensor-SVD~\cite{kilmer2013third,zhang2015denoising}. Furthermore learning sums of structured dictionaries can be used to represent tensor data~\cite{dantas2017learning,ghassemi2017stark}.

In this paper, our focus is on theoretical understanding of the fundamental limits of dictionary learning algorithms that explicitly account for the tensor structure of data in terms of \textit{Kronecker structured} (KS) dictionaries. It has been shown that many multidimensional signals can be decomposed into a superposition of separable atoms\cite{rivenson2009compressed,rivenson2009efficient,duarte2012kronecker}. In this case, a sequence of independent transformations on different data dimensions can be carried out using KS matrices. Such matrices have successfully been used for data representation in hyperspectral imaging, video acquisition, and distributed sensing\cite{duarte2012kronecker}.

To the best of our knowledge, none of the prior works on KS dictionary learning \cite{hawe2013separable,zubair2013tensor,roemer2014tensor,dantas2017learning} provide an understanding of the sample complexity of KS dictionary learning algorithms. In contrast, we provide lower bounds on the minimax risk of estimating KS dictionaries from tensor data using \emph{any} estimator. These bounds not only provide means of quantifying the performance of existing KS dictionary learning algorithms, but they also hint at the potential benefits of explicitly accounting for tensor structure of data during dictionary learning.

\subsection{Our Contributions}

Our first result is a general lower bound for the mean squared error (MSE) of estimating KS-dictionaries consisting of $K\geq 2$ coordinate dictionaries that sparsely represent $K$th-order tensor data. Here, we define the minimax risk to be the worst-case MSE that is attainable by the best dictionary estimator. Our approach uses the standard procedure for lower bounding the minimax risk in nonparametric estimation by connecting it to the maximum probability of error on a carefully constructed multiple hypothesis testing problem~\cite{tsybakov2009introduction,yu1997assouad}: the technical challenge is in constructing an appropriate set of hypotheses. In particular, consider a dictionary $\D \in \bbR^{m \times p}$ consisting of the Kronecker product of $K$ coordinate dictionaries $\D_k \in\bbR^{m_k\times p_k} , k \in \{1,\dots,K\}$, where $m=\prod_{k =1}^{K} m_k$ and $p=\prod_{k =1}^{K} p_k$, that is generated within the radius $r$ neighborhood (taking the Frobenius norm as the distance metric) of a fixed reference dictionary. Our analysis shows that given a sufficiently large $r$ and keeping some other parameters constant, a sample complexity\footnote{We use $f(n) = \mc{O}(g(n))$ and $f(n) = \Omega(g(n))$ if for sufficiently large $n \in \mathbb{N}$, $f(n)<C_1g(n)$ and $f(n)>C_2g(n)$, respectively, for some positive constants $C_1$ and $C_2$.} of $N=\Omega(\sum_{k=1}^{K}m_kp_k)$ is necessary for reconstruction of the true dictionary up to a given estimation error.
We also provide minimax bounds on the KS dictionary learning problem that hold for the following distributions for the coefficient vectors $\{\x_n\}$:
	\begin{itemize}
		\item 
		$\{\x_n\}$ are independent and identically distributed (i.i.d.) with zero mean and can have any distribution;
		\item 
		$\{\x_n\}$ are i.i.d. and sparse;
		\item 
		$\{\x_n\}$ are i.i.d., sparse, and their non-zero elements follow a Gaussian distribution.
	\end{itemize}

Our second contribution is development and analysis of an algorithm to learn dictionaries formed by the Kronecker product of 2 smaller dictionaries, which can be used to represent 2nd-order tensor data. To this end, we show that under certain conditions on the local neighborhood, the proposed algorithm can achieve one of the earlier obtained minimax lower bounds. 
Based on this, we believe that our lower bound may be tight more generally, but we leave this for future work.

\subsection{Relationship to Previous Work}

In terms of relation to prior work, theoretical insights into the problem of dictionary learning have either focused on specific algorithms for non-KS dictionaries~\cite{aharon2006uniqueness,agarwal2013learning,agarwal2013exact,arora2013new, schnass2014identifiability,schnass2014local,gribonval2014sparse} or lower bounds on minimax risk of dictionary learning for one-dimensional data~\cite{jung2014performance,jung2015minimax}. The former works provide sample complexity results for reliable dictionary estimation based on appropriate minimization criteria. Specifically, given a probabilistic model for sparse coefficients and a finite number of samples, these works find a local minimizer of a nonconvex objective function and show that this minimizer is a dictionary within a given distance of the true dictionary~\cite{schnass2014identifiability,schnass2014local,gribonval2014sparse}. In contrast, Jung et al.~\cite{jung2014performance,jung2015minimax} provide minimax lower bounds for dictionary learning from one-dimensional data under several coefficient vector distributions and discuss a regime where the bounds are tight in the scaling sense for some signal-to-noise ($\SNR$) values. In particular, for a given dictionary $\D$ and sufficiently large neighborhood radius $r$, they show that $N=\Omega(mp)$ samples are required for reliable recovery of the dictionary up to a prescribed MSE within its local neighborhood.
However, in the case of tensor data, their approach does not exploit the structure in the data, whereas our goal is to show how structure can potentially yield a lower sample complexity in the dictionary learning problem.

To provide lower bounds on the minimax risk of KS dictionary learning, we adopt the same general approach that Jung et al.~\cite{jung2014performance,jung2015minimax} use for the vector case. They use the standard approach of connecting the estimation problem to a multiple-hypothesis testing problem and invoking Fano's inequality~\cite{yu1997assouad}. We construct a family of KS dictionaries which induce similar observation distributions but have a minimum separation from each other.
By explicitly taking into account the Kronecker structure of the dictionaries, we show that the sample complexity satisfies a lower bound of $\Omega(\sum_{k=1}^{K}m_kp_k)$ compared to the $\Omega(mp)$ bound from vectorizing the data~\cite{jung2015minimax}.  Although our general approach is similar to that in~\cite{jung2015minimax}, there are fundamental differences in the construction of the KS dictionary class and analysis of the minimax risk.
This generalizes our preliminary work~\cite{shakeri2016minimax} from 2nd-order to $K$th-order and provides a comprehensive analysis of the KS dictionary class construction and minimax lower bounds.

Our results essentially show that the sample complexity depends linearly on the degrees of freedom of a Kronecker structured dictionary, which is $\sum_{k=1}^{K}m_kp_k$, and non-linearly  on the $\SNR$ and tensor order $K$. These lower bounds also depend on the radius of the local neighborhood around a fixed reference dictionary.
Our results hold even when some of the coordinate dictionaries are not overcomplete\footnote{Note that all coordinate dictionaries cannot be undercomplete, otherwise $\D$ won't be overcomplete.}. Like the previous work~\cite{jung2015minimax}, our analysis is local and our lower bounds depend on the distribution of multidimensional data.

We next introduce a KS dictionary learning algorithm for 2nd-order tensor data and show that in this case, one of the provided minimax lower bounds is achievable under certain conditions. We also conduct numerical experiments that demonstrate the empirical performance of the algorithm relative to the MSE upper bound and in comparison to the performance of a non-KS dictionary learning algorithm~\cite{jung2015minimax}.

\subsection{Notational Convention and Preliminaries}
\label{sec:notation}
Underlined bold upper-case, bold upper-case and lower-case letters are used to denote real-valued tensors, matrices and vectors, respectively. Lower-case letters denote scalars. The $k$-th column of $\X$ is denoted by $\x_k$ and its $ij$-th element is denoted by $x_{ij}$. Sometimes we use matrices indexed by multiple letters, such as $\X_{(a,b,c)}$, in which case its $j$-th column is denoted by $\x_{(a,b,c),j}$.  The function $\supp(.)$ denotes the locations of the nonzero entries of $\X$. Let $\X_{\mc{I}}$ be the matrix consisting of columns of $\X$ with indices $\mc{I}$, $\X^{\mc{T}}$ be the matrix consisting of rows of $\X$ with indices $\mc{T}$ and $\I_d$ be the $d\times d$ identity matrix. For a tensor $\uX \in \bbR^{p_1 \times \dots \times p_K}$, its $(i_1,\dots,i_K)$-th element is denoted as $\underline{x}_{i_1\dots i_K}$.
Norms are given by subscripts, so $\|\bu\|_0$ and $\|\bu\|_2$ are the $\ell_0$ and  $\ell_2$ norms of $\bu$, respectively, and $\|\X\|_2$ and $\|\X\|_F$ are the spectral and Frobenius norms of $\X$, respectively. We use $\vect(\X)$ to denote the vectorized version of matrix $\X$, which is a column vector obtained by stacking the columns of $\X$ on top of one another. We write $[K]$ for $\{1,\dots,K\}$. For matrices $\X$ and $\Y$, we define their distance in terms of the Frobenius norm:
	\begin{align*}
	d(\X,\Y)=\|\X-\Y\|_F.
	\end{align*}	
We define the outer product of two vectors of the same dimension, $\bu$ and $\bv$, as $\bu  \odot \bv = \bu \bv^\top$ and the inner product between matrices of the same size, $\X$ and $\Y$, as $\langle \X,\Y \rangle = \tr(\X^\top\Y)$. Furthermore, $P_{\mathcal{B}_1}(\bu)$ denotes the projection of $\bu$ on the closed unit ball, i.e.,
	\begin{align}
	P_{\mathcal{B}_1}(\bu) =
	\begin{cases}
		\bu   ,                \quad \quad \text{if} \ \|\bu\|_2 \leq 1, \\
		\frac{\bu}{\|\bu\|_2} ,\quad \text{otherwise}.
	\end{cases}
	\end{align}	

We now define some important matrix products.
We write $\X \otimes \Y$ for the \textit{Kronecker product} of two matrices $\X\in \bbR^{m\times n}$ and $\Y\in \bbR^{p\times q}$, defined as
	\begin{align}
	\X \otimes \Y =
		\begin{bmatrix}
		x_{11} \Y & x_{12} \Y & \dots & x_{1n} \Y \\
		\vdots & \vdots & \ddots & \vdots\\
		x_{m1} \Y & x_{m2} \Y & \dots & x_{mn} \Y \\
		\end{bmatrix},
	\end{align}
where the result is an $mp \times nq$ matrix and we have $\|\X \otimes\Y \|_F = \|\X\|_F\| \Y\|_F$~\cite{horn2012matrix}. Given matrices $\X_1, \X_2, \Y_1$, and $\Y_2$, where products $\X_1\Y_1$ and $\X_2\Y_2$ can be formed, we have~\cite{smilde2005multi}
	\begin{align} \label{eq:Kron_prod}
	(\X_1 \otimes \X_2)(\Y_1 \otimes \Y_2) = (\X_1 \Y_1) \otimes (\X_2 \Y_2).
	\end{align}
Given $\X\in \bbR^{m\times n}$ and $\Y\in \bbR^{p\times n}$, we write $\X \ast \Y$ for their $mp \times n$ \textit{Khatri-Rao product}~\cite{smilde2005multi}, defined by
	\begin{align}
	\X \ast \Y =
		\begin{bmatrix}
		\x_1 \otimes \y_1 & \x_2 \otimes \y_2 & \dots & \x_n \otimes \y_n
		\end{bmatrix}.
	\end{align}
This is essentially the column-wise Kronecker product of matrices $\X$ and $\Y$. We also use $\bigotimes_{k \in K} \X_k = \X_1 \otimes \dots \otimes \X_K$ and $\bigAst_{k \in K} \X_k = \X_1 \ast \dots \ast \X_K$.

Next, we review essential properties of $K$th-order tensors and the relation between tensors and the Kronecker product of matrices using the \textit{Tucker decomposition} of tensors.

\subsubsection{A Brief Review of Tensors}

A tensor is a multidimensional array where the order of the tensor is defined as the number of components in the array.
A tensor $\uX \in \bbR^{p_1 \times p_2\times \dots \times p_K}$ of order $K$ can be expressed as a matrix by reordering its elements to form a matrix. This reordering is called unfolding: the mode-$k$ unfolding matrix of a tensor is a $p_k \times \prod_{i \ne k} p_i$ matrix, which we denote by $\X_{(k)}$. Each column of $\X_{(k)}$ consists of the vector formed by fixing all indices of $\uX$ except the one in the $k$th-order. For example, for a 2nd-order tensor $\uX$, the mode-1 and mode-2 unfolding matrices are $\uX$ and $\uX^\top$, respectively.
The $k$-rank of a tensor $\uX$ is defined by $\rnk(\X_{(k)})$; trivially, $\rnk(\X_{(k)}) \leq p_k$.

The mode-$k$ matrix product of the tensor $\uX$ and a matrix $\A \in \bbR^{m_k \times p_k}$, denoted by $\uX \times_k \A$, is a tensor of size $p_1 \times \dots p_{k-1} \times m_k \times p_{k+1} \dots \times p_K$ whose elements are
	\begin{align}
	(\uX \times_k \A)_{i_1\dots i_{k-1} j i_{k+1} \dots i_K} = \sum_{i_k=1}^{p_k} \underline{x}_{i_1\dots i_{k-1} i_k i_{k+1} \dots i_K} a_{ji_k}.
	\end{align}
The mode-$k$ matrix product of $\uX$ and $\A$ and the matrix multiplication of $\X_{(k)}$ and $\A$ are related~\cite{kolda2009tensor}:
	\begin{align}
	\uY = \uX \times_k \A \Leftrightarrow \Y_{(k)} = \A \X_{(k)}.
	\end{align}

\subsubsection{Tucker Decomposition for Tensors}
The Tucker decomposition is a powerful tool that decomposes a tensor into a \textit{core tensor} multiplied by a matrix along each mode~\cite{tucker1963implications,kolda2009tensor}. We take advantage of the Tucker model since we can relate the Tucker decomposition to the Kronecker representation of tensors~\cite{caiafa2013computing}.
For the tensor $\uY \in \bbR^{m_1 \times m_2 \times \dots \times m_K}$ of order $K$, if $\rnk(\Y_{(k)})\leq p_k$ holds for all $k \in [K]$ then, according to the Tucker model, $\uY$ can be decomposed into:
	\begin{align} \label{eq:UY_UX}
	\uY = \uX \times_1 \D_1  \times_2 \D_2 \times_ 3 \dots \times_K \D_K,
	\end{align}
where $\uX \in \bbR^{p_1 \times p_2\times \dots \times p_K}$ denotes the core tensor and $\D_k \in \bbR^{m_k \times p_k}$ are factor matrices. Here, \eqref{eq:UY_UX} can be interpreted as a form of higher order principal component analysis (PCA):
	\begin{align}
	\uY = \sum_{i_1 \in [p_1]} \dots \sum_{i_K \in [p_K]}
		\underline{x}_{i_1\dots i_K} \bd_{1,i_1} \odot \dots \odot \bd_{K,i_K},
	\end{align}
where the $\D_k$'s can be interpreted as the principal components in mode-$k$.
The following is implied by \eqref{eq:UY_UX}~\cite{kolda2009tensor}:
	\begin{align}
	\Y_{(k)} = \D_{k}\X_{(k)}(\D_{K} \otimes \dots \otimes \D_{k+1} \otimes \D_{k-1} \otimes \dots \otimes \D_1)^\top.
	\end{align}
Since the Kronecker product satisfies $\vect(\B\X\A^\top)=(\A \otimes \B)\vect(\X)$, \eqref{eq:UY_UX} is equivalent to
	\begin{align} \label{eq:vecty_vectx}
	\vect(\uY) = \big( \D_K \otimes \D_{K-1} \otimes \dots \otimes \D_1 \big) \vect(\uX),
	\end{align}
where $\vect(\uY) \triangleq \vect(\Y_{(1)})$ and $\vect(\uX) \triangleq \vect(\X_{(1)})$~ \cite{van2000ubiquitous,kolda2009tensor,caiafa2013computing}.


The rest of the paper is organized as follows. We formulate the KS dictionary learning problem and describe the procedure for obtaining minimax risk lower bounds in Section~\ref{sec:SysM}. Next, we provide a lower bound for general coefficient distribution in Section~\ref{sec:LB_General} and in Section~\ref{sec:LB_Sparse}, we present lower bounds for sparse and sparse Gaussian coefficient vectors. We propose a KS dictionary learning algorithm for 2nd-order tensor data and analyze its corresponding MSE and empirical performance in Section~\ref{sec:prtl_cnvrse}. In Section~\ref{sec:discussion}, we discuss and interpret the results. Finally, in Section~\ref{sec:conclusion}, we conclude the paper. In order to keep the main exposition simple, proofs of most of the lemmas and theorems are relegated to the appendix.

\section{Problem Formulation}
\label{sec:SysM}
In the conventional dictionary learning model, it is assumed that the observations $\y_n \in \bbR^m$ are generated via a fixed dictionary as
	\begin{align}
	\y_n= \D \x_n + \boldsymbol{\eta}_n,
	\end{align}
in which the dictionary $\D\in \bbR^{m\times p}$ is an overcomplete basis ($m<p$) with unit-norm columns\footnote{The unit-norm condition on columns of $\D$ is required to avoid solutions with arbitrary large norms for dictionary columns and  small values for $\X$.} and rank $m$, $\x_n \in \bbR^p$ is the coefficient vector, and $\boldsymbol{\eta}_n \in \bbR^m$ denotes observation noise.

Our focus in this work is on multidimensional signals. We assume the observations are $K$th-order tensors $\uY_n \in \bbR^{m_1\times m_2 \times \dots \times m_K}$. According to the Tucker model, given \textit{coordinate dictionaries} $\D_k \in \bbR^{m_k \times p_k}$, a \textit{coefficient tensor} $\uX_n \in \bbR^{p_1\times p_2 \times \dots \times p_K}$, and a \textit{noise tensor} $\uN_n$, we can write $\y_n \triangleq \vect(\uY_n)$ using \eqref{eq:vecty_vectx} as\footnote{We have reindexed $\D_k$'s in \eqref{eq:vecty_vectx} for ease of notation.}
	\begin{align} \label{eq:model}
	\y_n= \bigg( \bigotimes_{k \in [K]} \D_k \bigg) \x_n + \boldsymbol{\eta}_n,
	\end{align}
where $\x_n \triangleq \vect(\uX_n)$ and $\boldsymbol{\eta}_n \triangleq \vect(\uN_n)$. Let
	\begin{align}
	m = \prod_{k \in [K]} m_k \quad \text{and} \quad
		p = \prod_{k \in [K]} p_k.
	\end{align}	
Concatenating $N$ i.i.d. noisy observations $\{\y_n\}_{n=1}^N$, which are realizations according to the model \eqref{eq:model}, into $\Y\in \bbR^{m\times N}$, we obtain
	\begin{align} \label{eq:Y}
	\Y=\D\X+\N,
	\end{align}
where $\D \triangleq \bigotimes_{k \in [K]} \D_k$ is the unknown KS dictionary, $\X \in \bbR^{p\times N} $ is a coefficient matrix consisting of i.i.d. random coefficient vectors with known distribution that has zero-mean and covariance matrix $\Sig_x$, and $\N\in \bbR^{m\times N}$ is assumed to be additive white Gaussian noise (AWGN) with zero mean and variance $\sigma^2$.
	
Our main goal in this paper is to derive necessary conditions under which the KS dictionary $\D$ can possibly be learned from the noisy observations given in \eqref{eq:Y}. We assume the true KS dictionary $\D$ consists of unit-norm columns and we carry out local analysis. That is, the true KS dictionary $\D$ is assumed to belong to a neighborhood around a fixed (normalized) reference KS dictionary
	\begin{align} \label{eq:D_0}
	\D_0 = \bigotimes_{k \in [K]} \D_{(0,k)} ,
	\end{align}	
and $\D_0 \in \cD$, where
	\begin{align}
	\cD \triangleq  & \bigg\{  \D' \in \bbR^{m\times p} : \ \D'= \bigotimes_{k \in [K]} \D_k',
		\D'_k \in \bbR^{m_k \times p_k}, \nonumber \\
	&\qquad\|\bd'_{k,j}\|_2=1 \ \forall k \in [K], j\in [p_k] \
		 \bigg\}. \label{eq:D0}
	\end{align}
We assume the true generating KS dictionary $\D$ belongs to a neighborhood around $\D_0$:
	\begin{align}
	\D \in \cX(\D_0,r)  \triangleq  \lr{ \D' \in \cD : \norm{\D' -\D_0}_F < r} \label{Dclass}
	\end{align}
for some fixed radius $r$.\footnote{Note that our results hold with the unit-norm condition enforced only on $\D$ itself, and not on the subdictionaries $\D_k$. Nevertheless, we include this condition in the dictionary class for the sake of completeness as it also ensures uniqueness of the subdictionaries (factors of a $K$-fold Kronecker product can exchange scalars $\gamma_k$ freely without changing the product as long as $\prod_{k\in[K]} \gamma_k = 1$).} Note that $\D_0$ appears in the analysis as an artifact of our proof technique to construct the dictionary class. In particular, if $r$ is sufficiently large, then $\cX(\D_0,r) \approx \cD$ and effectively $\D \in \cD$.

\subsection{Minimax Risk}

We are interested in lower bounding the minimax risk for estimating $\D$ based on observations $\Y$, which is defined as the worst-case mean squared error (MSE) that can be obtained by the best KS dictionary estimator $\wh{\D}(\Y)$. That is,
\begin{align} \label{eq:minimax}
	\eps^* = \inf_{\wh{\D}} \sup_{\D\in \cX(\D_0,r)} \bbE_{\Y} \lr{ \big\|\wh{\D}(\Y)-\D\big\|_F^2},
	\end{align}
where $\wh{\D}(\Y)$ can be estimated using any KS dictionary learning algorithm.
In order to lower bound this minimax risk $\eps^*$, we employ a standard reduction to the multiple hypothesis testing used in the literature on nonparametric estimation~\cite{yu1997assouad,tsybakov2009introduction}.  This approach is equivalent to generating a KS dictionary $\D_l$ uniformly at random from a carefully constructed class $\cD_L=\{\D_1, \dots, \D_L\} \subseteq  \cX(\D_0,r), L \geq 2,$ for a given $(\D_0$, $r)$.
To ensure a tight lower bound, we must construct $\cD_L$ such that the distance between any two dictionaries in $\cD_L$ is large but the hypothesis testing problem is hard; that is, two distinct dictionaries $\D_l$ and $\D_{l'}$ should produce similar observations. Specifically, for $l,l' \in [L]$, and given error $\eps \geq \eps^*$, we desire a construction such that
	\begin{align}
	\forall l \not= l', &\norm{\D_l - \D_{l'} }_F \geq 2\sqrt{\gamma\eps}\quad \text{and}\nonumber\\
	&D_{KL} \left( f_{\D_l}(\Y)|| f_{\D_{l'}}(\Y) \right) \leq \alpha_L,\label{eqn:minimax}
	\end{align}
where $D_{KL} \left( f_{\D_l}(\Y)|| f_{\D_{l'}}(\Y) \right)$ denotes the Kullback-Leibler (KL) divergence between the distributions of observations based on $\D_l \in \cD_L$ and $\D_{l'} \in \cD_L$, while $\gamma$, $\al_L$, and $\eps$ are non-negative parameters.  Observations $\Y=\D_l\X+\N$ in this setting can be interpreted as channel outputs that are used to estimate the input $\D_l$ using an arbitrary KS dictionary algorithm that is assumed to achieve the error $\eps$. Our goal is to detect the correct generating KS dictionary index $l$. For this purpose, a minimum distance detector is used:
	\begin{align}
	\wh{l} = \min_{l' \in [L]}  \left\|\wh{\D}(\Y) - \D_{l'}\right\|_F.
	\end{align}	
Then, we have $\bbP (\wh{l}(\Y)\neq l) = 0$ for the minimum-distance detector $\wh{l}(\Y)$ as long as $\|\wh{\D}(\Y)-\D_l\|_F < \sqrt{\gamma\eps}$. The goal then is to relate $\eps$ to $\bbP(\|\wh{\D}(\Y)-\D_l\|_F \geq \sqrt{\gamma\eps})$ and $\bbP (\wh{l}(\Y)\neq l)$ using Fano's inequality~\cite{yu1997assouad}:
	\begin{align}
    \label{eqn:fano}
	(1- \bbP (\wh{l}(\Y)\neq l)) \log_2 L - 1 \leq I(\Y;l),
	\end{align}
where $I(\Y;l)$ denotes the mutual information (MI) between the observations $\Y$ and the dictionary $\D_l$. Notice that the smaller $\alpha_L$ is in \eqref{eqn:minimax}, the smaller $I(\Y;l)$ will be in \eqref{eqn:fano}. Unfortunately, explicitly evaluating $I(\Y;l)$ is a challenging task in our setup because the underlying distributions are mixture of distributions. Similar to~\cite{jung2015minimax}, we will instead resort to upper bounding $I(\Y;l)$ by conditioning it on some side information $\T(\X)$ that will make the observations $\Y$ conditionally multivariate Gaussian (in particular, from~\cite[Lemma A.1]{jung2015minimax}, it follows that $I(\Y;l) \leq I(\Y;l|\T(\X))$).\footnote{Instead of upper bounding $I(\Y;l|\T(\X))$, similar results can be derived by using Fano's inequality for the conditional probability of error, $\bbP(\wh{l}(\Y)\neq l|\T(\X))$\cite[Theorem 2]{wainwright2009information}. }
We will in particular focus on two types of side information: $\T(\X)=\X$ and $\T(\X) = \supp(\X)$. 
A lower bound on the minimax risk in this setting depends not only on problem parameters such as the number of observations $N$, noise variance $\sigma^2$, dimensions $\{m_k\}_{k=1}^K$ and $\{p_k\}_{k=1}^K$ of the true KS dictionary, neighborhood radius $r$, and coefficient covariance $\Sig_x$, but also on the structure of the constructed class $\cD_L$~\cite{tsybakov2009introduction}.
Note that our approach is applicable to the global KS dictionary learning problem, since the minimax lower bounds that are obtained for any $\D \in \cX(\D_0,r)$ are also trivially lower bounds for $\D \in \cD$.

After providing minimax lower bounds for the KS dictionary learning problem, we develop and analyze a simple KS dictionary learning algorithm for $K=2$ order tensor data. Our analysis shows that one of our provided lower bounds is achievable, suggesting that they may be tight.

\subsection{Coefficient Distribution}

By making different assumptions on coefficient distributions, we can specialize our lower bounds to specific cases. To facilitate comparisons with prior work, we adopt somewhat similar coefficient distributions as in the unstructured case~\cite{jung2015minimax}. First, we consider any coefficient distribution and only assume that the coefficient covariance matrix exists. We then specialize our analysis to sparse coefficient vectors and, by adding additional conditions on the reference dictionary $\D_0$, we obtain a tighter lower bound for the minimax risk for some SNR regimes.

\subsubsection{General Coefficients}

First, we consider the general case, where $\x$ is a zero-mean random coefficient vector with covariance matrix $\Sig_x = \bbE_\x \lr{\x \x^\top}$. We make no additional assumption on the distribution of $\x$. We condition on side information $\T(\X)=\X$ to obtain a lower bound on the minimax risk in the case of general coefficients.

\subsubsection{Sparse Coefficients}

In the case where the coefficient vector is sparse, we show that additional assumptions on the non-zero entries yield a lower bound on the minimax risk conditioned on side information $\supp(\x)$, which denotes the support of $\x$ (the set containing indices of the locations of the nonzero entries of $\x$). We study two cases for the distribution of $\supp(\x)$:

\begin{itemize}
\item \textbf{Random Sparsity.}
In this case, the random support of $\x$ is distributed uniformly over $\cE_1=\{\cS\subseteq [p]:|\cS|=s\}$:
	\begin{align} \label{swiss}
	\bbP(\supp(\x)=\cS)=\frac{1}{
	{p \choose s}},
	\quad \text{for any} \ \cS \in \cE_1.
\end{align}

\item \textbf{Separable Sparsity.} In this case we sample $s_k$ elements uniformly at random from $[p_k]$, for all $k \in [K]$. The random support of $\x$ is $\cE_2=\{\cS \subseteq [p]:|\cS|=s\}$, where $\cS$ is related to $\{\cS_1 \times \dots \times \cS_K: \cS_k\subseteq [p_k], |\cS_k|=s_k, k \in [K]\}$ via lexicographic indexing. The number of non-zero elements in $\x$ in this case is $s=\prod_{k \in [K]} s_k$. The probability of sampling $K$ subsets $\{\mc{S}_1, \dots, \mc{S}_K\} $ is
	\begin{align} \label{crack}
	\bbP(\supp(\x)=\cS)=
	\frac{1}{ \prod_{k \in [K]}
	{p_k \choose s_k}},
	\quad \text{for any} \ \cS \in \cE_2.
	\end{align}
\end{itemize}

In other words, separable sparsity requires non-zero coefficients to be grouped in blocks. This model arises in the case of processing of images and video sequences~\cite{caiafa2013computing}.

\begin{remark} If $\uX$ follows the separable sparsity model with sparsity $(s_1,\dots,s_K)$, then the columns of the mode-$k$ matrix $\Y_{(k)}$ of $\uY$ have $s_k$-sparse representations with respect to $\D_k$, for $k \in [K]$~\cite{caiafa2013computing}.
\end{remark}

For a signal $\x$ with sparsity pattern $\supp(\x)$, we model the non-zero entries of $\x$, i.e., $\x_\cS$, as drawn independently and identically from a probability distribution with known variance $\sigma_a^2$:
	\begin{align} \label{iid}
	\bbE_x\{\x_\cS\x_\cS^T|\cS\}=\sigma_a^2 \I_s.
	\end{align}
Any $\x$ with sparsity model \eqref{swiss} or \eqref{crack} and nonzero entries satisfying \eqref{iid} has covariance matrix
	\begin{align} \label{sig_iid}
	\Sig_x = \frac{s}{p}\sigma_a^2 \I_p.
	\end{align}

\section{Lower Bound for General Distribution}
\label{sec:LB_General}

We now provide our main result for the lower bound for minimax risk of the KS dictionary learning problem for the case of general coefficient distributions.

\begin{theorem}\label{thm_1}
Consider a KS dictionary learning problem with $N$ i.i.d. observations generated according to model~\eqref{eq:model}. Suppose the true dictionary satisfies \eqref{Dclass} for some $r$ and fixed reference dictionary $\D_0$ satisfying \eqref{eq:D_0}. Then for any coefficient distribution with mean zero and covariance $\Sig_x$, we have the following lower bound on $\eps^*$:
	\begin{align} \label{eq:thm_1}
	\eps^* \geq \frac{t}{4}  \min \bigg\{
		p , \frac{r^2}{2K} ,  &
		\frac{\sigma^2}{4NK\|\Sig_x\|_2}
		\bigg(c_1\bigg(\sum_{k \in [K]} (m_k-1)p_k\bigg) \nonumber \\
 	&\qquad \qquad -\frac{K}{2}\log_2 2K -2\bigg)\bigg\},
	\end{align}
for any $0<t<1$ and any $0<c_1<\dfrac{1-t}{8\log 2}$.
\end{theorem}

The implications of Theorem~\ref{thm_1} are examined in Section~\ref{sec:discussion}.

\textit{Outline of Proof:} The idea of the proof is that we construct a set of $L$ distinct KS dictionaries, $\cD_L =\{\D_1,\dots,\D_L\} \subset \cX(\D_0,r)$, such that any two distinct dictionaries are separated by a minimum distance. That is for any pair $l,l' \in [L]$ and any positive $\eps<\dfrac{tp}{4} \min\left\{r^2,\dfrac{r^4}{2Kp}\right\}$:
	\begin{align}
	\label{eq:distance}
	\|\D_l-\D_l'\|_F \geq 2\sqrt{2\eps}, \ \text{for} \ l \neq l'.
	\end{align}
In this case, if a dictionary $\D_l \in \cD_L$ is selected uniformly at random from $\cD_L$, then conditioned on side information $\T(\X)=\X$, the observations under this dictionary follow a multivariate Gaussian distribution. We can therefore upper bound the conditional MI by approximating the upper bound for KL-divergence of multivariate Gaussian distributions. This bound depends on parameters $\eps, N, \{m_k\}_{k=1}^K, \{p_k\}_{k=1}^K, \Sig_x, s, r, K$, and $\sigma^2$.

Assuming \eqref{eq:distance} holds for $\cD_L$, if there exists an estimator achieving the minimax risk $\eps^*\leq \eps$ and the recovered dictionary $\wh{\D}(\Y)$ satisfies $\|\wh{\D}(\Y)-\D_l\|_F < \sqrt{2\eps}$, the minimum distance detector can recover $\D_l$. Then, using the Markov inequality and since $\eps^*$ is bounded, the probability of error $\bbP(\wh{\D}(\Y) \neq \D_l) \leq \bbP(\|\wh{\D}(\Y)-\D_l\|_F \geq \sqrt{2\eps})$ can be upper bounded by $\frac{1}{2}$. Further, according to \eqref{eqn:fano}, the lower bound for the conditional MI can be obtained using Fano's inequality~\cite{jung2015minimax}. The  lower bound is a function of $L$ only.
Finally, using the obtained bounds for the conditional MI, we derive a lower bound for the minimax risk $\eps^*$.

\begin{remark}
We use the constraint in \eqref{eq:distance} in Theorem~\ref{thm_1} for simplicity: the number $2\sqrt{2}$ can be replaced with any arbitrary $\gamma>0$.
\end{remark}

The complete technical proof of Theorem \ref{thm_1} relies on the following lemmas, which are formally proved in the appendix.
Although the similarity of our model to that of Jung et al.~\cite{jung2015minimax} suggests that our proof should be a simple extension of their proof of Theorem 1, the construction for KS dictionaries is more complex and its analysis requires a different approach. One exception is Lemma \ref{lemma_3}~\cite[Lemma 8]{jung2015minimax}, which connects a lower bound on the Frobenius norms of pairwise differences in the construction to a lower bound on the conditional MI used in Fano's inequality~\cite{yu1997assouad}.

\begin{lemma} \label{lemma_McD}
Let $\alpha > 0$ and $\beta > 0$. Let $\{ \A_l \in \bbR^{m\times p} : l \in [L] \}$ be a set of $L$ matrices  where each $\A_l$ contains $m \times p$ independent and identically distributed random variables taking values $\pm \alpha$ uniformly. Then we have the following inequality:
	\begin{align} \label{eq:lemma_McD}
	\bbP\left(\exists (l,l') \in [L]\times [L], l \neq l': \left|\ip{\A_l}{ \A_{l'}}\right| \geq \beta\right) \nonumber \\
	\leq 2L^2\exp\left(-\frac{\beta^2}{4\alpha^4mp} \right).
	\end{align}
\end{lemma}

\begin{lemma}\label{lemma_2}
Consider the generative model in \eqref{eq:model}. Fix $r>0$ and a reference dictionary $\D_0$ satisfying \eqref{eq:D_0}. Then there exists a set $\cD_L\subseteq \cX (\D_0,r)$ of cardinality $L=2^{\flr{ c_1(\sum_{k \in[K]}(m_k-1)p_k)-\frac{K}{2}\log_2 (2K)}}$ such that for any $0<t<1$, any $0<c_1 < \frac{t^2}{8\log 2}$, any $\eps'>0$ satisfying
	\begin{align}
	\eps' &< r^2 \min  \left\{ 1, \frac{r^2}{2Kp} \right\}, \label{eq:eps_r}
	\end{align}
and all pairs $l,l' \in [L]$, with $l \neq l'$, we have
	\begin{align}
	\frac{2p}{r^2}(1-t) \eps' &\leq \|\D_l-\D_{l'}\|_F^2  \leq \frac{4Kp}{r^2}\eps'.
	\end{align}
Furthermore, if $\X$ is drawn from a distribution with mean $\mathbf{0}$ and covariance matrix $\Sig_x$ and conditioning on side information $\T(\X)=\X$, we have
	\begin{align} \label{eq:I_1}
	I(\Y;l|\T(\X)) &\leq \frac{2NKp\|\Sig_x\|_2}{r^2\sigma^2}\eps'.
	\end{align}
\end{lemma}

\begin{lemma}[Lemma 8~\cite{jung2015minimax}] \label{lemma_3} Consider the generative model in \eqref{eq:model} and suppose the minimax risk $\eps^*$ satisfies $\eps^* \leq \eps$ for some $\eps>0$. If there exists a finite set $\cD_L \subseteq \cD$ with $L$ dictionaries satisfying
	\begin{align} \label{eq:lem3_cons}
	\|\D_l-\D_{l'}\|_F^2\geq 8\eps
	\end{align}
for $l \neq l'$, then for any side information $\T(\X)$, we have
	\begin{align}
	I(\Y;l|\T(\X)) \geq \frac{1}{2} \log_2 (L) -1. \label{LB}
	\end{align}
\end{lemma}

\begin{IEEEproof}[Proof of Lemma~\ref{lemma_3}]
The proof of Lemma~\ref{lemma_3} is identical to the proof of Lemma 8 in Jung et al.~\cite{jung2015minimax}.
\end{IEEEproof}

\begin{IEEEproof}[Proof of Theorem \ref{thm_1}]
According to Lemma~\ref{lemma_2}, for any $\eps'$ satisfying \eqref{eq:eps_r}, there exists a set $\cD_L \subseteq \cX(\D_0,r)$ of cardinality $L=2^{\flr{ c_1(\sum_{k\in[K]} (m_k-1)p_k)-\frac{K}{2}\log_2 (2K) }}$ that satisfies \eqref{eq:I_1} for any $0<t'<1$ and any $c_1<\dfrac{t'}{8\log 2}$ . Let $t=1-t'$. If there exists an estimator with worst-case MSE satisfying $\eps^*\leq \dfrac{2tp}{8} \min\left\{1,\dfrac{r^2}{2Kp}\right\}$ then, according to Lemma \ref{lemma_3}, if we set $\frac{2tp}{r^2}\eps' = 8\eps^*$, \eqref{eq:lem3_cons} is satisfied for $\cD_L$ and \eqref{LB} holds.
Combining \eqref{eq:I_1} and \eqref{LB} we get
	\begin{align} \label{eq:dovar}
	\frac{1}{2} \log_2 (L) -1 \leq I(\Y;l|\T(\X))
		\leq  \frac{16 NKp\|\Sig_x\|_2}{c_2r^2\sigma^2}\eps^*,
	\end{align}
where $c_2 =\dfrac{2tp}{r^2}$. We can write \eqref{eq:dovar} as
	\begin{align}
	\eps^* \geq \frac{t\sigma^2}{16NK\|\Sig_x\|_2}
		 \bigg(c_1\bigg(\sum_{k\in[K]} (m_k-1)p_k\bigg) \nonumber \\
		 -\frac{K}{2}\log_2 2K-2\bigg). \label{thm_1_eps}
	\end{align}
\end{IEEEproof}

\section{Lower Bound for Sparse Distributions}
\label{sec:LB_Sparse}

We now turn our attention to the case of sparse coefficients and obtain lower bounds for the corresponding minimax risk. We first state a corollary of Theorem \ref{thm_1} for sparse coefficients, corresponding to $\T(\X)=\X$.

\begin{cor} \label{cor_1}
Consider a KS dictionary learning problem with $N$ i.i.d. observations generated according to model~\eqref{eq:model}. Suppose the true dictionary satisfies \eqref{Dclass} for some $r$ and fixed reference dictionary $\D_{0}$ satisfying \eqref{eq:D_0}. If the random coefficient vector  $\x$ is selected according to \eqref{swiss} or \eqref{crack}, we have the following lower bound on $\eps^*$:
	\begin{align}\label{eq:cor_1}
	\eps^* &\geq \frac{t}{4}  \min \bigg\{
		p ,\frac{r^2}{2K},
		\frac{\sigma^2 p}{4 N K s\sigma_a^2}
		\bigg(c_1 \bigg(\sum_{k \in [K]}(m_k-1)p_k\bigg)
		\nonumber \\
	&\qquad \qquad \qquad \qquad \qquad \ \qquad
		 -\frac{K}{2}\log_2 2K - 2 \bigg) \bigg\},
	\end{align}
for any $0<t<1$ and any $0<c_1<\dfrac{1-t}{8\log 2}$.
\end{cor}
This result is a direct consequence of Theorem \ref{thm_1}, obtained by substituting the covariance matrix of sparse coefficients given in \eqref{sig_iid} into \eqref{eq:thm_1}.

\subsection{Sparse Gaussian Coefficients}

In this section, we make an additional assumption on the coefficient vectors generated according to \eqref{swiss} and assume non-zero elements of the vectors follow a Gaussian distribution.
By additionally assuming the non-zero entries of $\x$ are i.i.d. Gaussian distributed, we can write $\x_\cS$ as
\begin{align} \label{gaussian}
\x_\cS \sim \mc{N}(\mb{0},\sigma_a^2\I_s).
\end{align}
As a result, conditioned on side information $\T(\x_n)=\supp(\x_n)$, observations $\y_n$ follow a multivariate Gaussian distribution. Part of our forthcoming analysis relies on the notion of the \emph{restricted isometry property} ($\RIP$) for a matrix.
\subsubsection*{Restricted Isometry Property ($\RIP$)\cite{candes2005decoding}}
A matrix $\wt{\D}$ with unit $\ell_2$-norm columns satisfies the $\RIP$ of order $s$ with constant $\delta_s$ if
\begin{align}
(1-\delta_s)\|\x\|_2^2 \leq \|\wt{\D}\x\|_2^2 \leq (1+\delta_s)\|\x\|_2^2,
\end{align}
for all $\x$ such that $\|\x\|_0 \leq s$.

We now provide a lower bound on the minimax risk in the case of coefficients selected according to \eqref{swiss} and \eqref{gaussian}.

\begin{theorem} \label{thm_2}
Consider a KS dictionary learning problem with $N$ i.i.d. observations generated according to model~\eqref{eq:model}. Suppose the true dictionary satisfies \eqref{Dclass} for some $r$ and fixed reference dictionary satisfying \eqref{eq:D_0}. If the reference coordinate dictionaries $\{\D_{0,k}, k \in [K]\}$ satisfy $\RIP(s,\frac{1}{2})$ and the random coefficient vector $\x$ is selected according to \eqref{swiss} and \eqref{gaussian}, we have the following lower bound on $\eps^*$:
	\begin{align} \label{eq:thm_2}
	\eps^* &\geq \frac{t}{4} \min \bigg\{
	    \frac{p}{s} ,\frac{r^2}{2K},
		\frac{\sigma^4p}{36(3^{4K})Ns^2 \sigma_a^4} \nonumber \\
	& \qquad \
		\bigg(c_1\bigg(\sum_{k \in [K]}(m_k-1) p_k\bigg)
		 -\frac{1}{2}\log_2 2K-2\bigg)\bigg\},
	\end{align}
for any $0<t<1$ and any $0<c_1<\dfrac{1-t}{8\log 2}$.
\end{theorem}

Note that in Theorem~\ref{thm_2}, $\D$ (or its coordinate dictionaries) need not satisfy the $\RIP$ condition. Rather, the $\RIP$ is only needed for the coordinate reference dictionaries, $\{\D_{0,k}, k \in [K]\}$, which is a significantly weaker (and possibly trivial to satisfy) condition. We state a variation of Lemma~\ref{lemma_2} necessary for the proof of Theorem~\ref{thm_2} --- the proof is provided in the appendix.

\begin{lemma}\label{lemma_Sp_I_UB} Consider the generative model in \eqref{eq:model}. Fix $r>0$ and reference dictionary $\D_0$ satisfying \eqref{eq:D_0}. Then, there exists a set $\cD_L\subseteq \cX(\D_0,r)$ of cardinality $L=2^{\flr{ c_1(\sum_{k \in [K]}(m_k-1)p_k)-\frac{1}{2}\log_2(2K)}}$ such that for any $0<t<1$, any $0<c_1<\frac{t^2}{8\log 2}$, any $\eps'>0$ satisfying
	\begin{align} \label{eps:2}
	0<\eps'\leq r^2 \min\left\{\frac{1}{s},\frac{r^2}{2Kp}\right\},
	\end{align}
and any $l,l' \in [L]$, with $l \neq l'$, we have
\begin{align} \label{eq:8eps_Sprs}
\frac{2p}{r^2}(1-t)\eps' \leq \|\D_l-\D_{l'}\|_F^2\leq \frac{4Kp}{r^2}\eps'.
\end{align}
Furthermore, assuming the reference coordinate dictionaries $\{\D_{0,k}, k \in [K]\}$ satisfy $\RIP(s,\frac{1}{2})$, the coefficient matrix $\X$ is selected according to \eqref{swiss} and \eqref{gaussian}, and considering side information $\T(\X)=\supp(\X)$, we have:
	\begin{align} \label{eq:I_2}
	I(\Y;l|\T(\X))&\leq 36( 3^{4K}  ) \left(\frac{\sigma_a}{\sigma}\right)^4 \frac{Ns^2}{r^2}\eps'.
	\end{align}
\end{lemma}

\begin{IEEEproof} [ Proof of Theorem \ref{thm_2}]
According to Lemma \ref{lemma_Sp_I_UB}, for any $\eps'$ satisfying \eqref{eps:2}, there exists a set $\cD_L \subseteq \cX(\D_0,r)$ of cardinality $L=2^{\flr{ c_1(\sum_{k \in [K]}(m_k-1)p_k)-\frac{K}{2}\log_2(2K)}}$ that satisfies \eqref{eq:I_2} for any $0<t'<1$ and any $c_1<\frac{t'}{8\log 2}$. Denoting $t=1-t'$ and provided there exists an estimator with worst case MSE satisfying $\eps^*\leq \dfrac{tp}{4} \min\big\{\dfrac{1}{s},\dfrac{r^2}{2Kp}\big\}$, if we set $\dfrac{2tp}{r^2}\eps'=8\eps^*$, \eqref{eq:lem3_cons} is satisfied for $\cD_L$ and  \eqref{LB} holds. Consequently,
	\begin{align} \label{dovar2}
	\frac{1}{2} \log_2 (L) -1 \leq I(\Y;l|\T(\X))
		\leq  \frac{36(3^{4K})}{c_2} \left(\frac{\sigma_a}{\sigma}\right)^4\frac{Ns^2}{r^2} \eps^*,
	\end{align}
where $c_2 = \dfrac{p(1-t)}{4r^2}$. We can write \eqref{dovar2} as
	\begin{align} \label{thm_2.1_eps}
	\eps^* \geq \big(\frac{\sigma}{\sigma_a}\big)^4\frac{ tp \left(c_1\left(\sum_{k \in [K]} (m_k-1)p_k\right) -\frac{K}{2}\log_2 2K - 2 \right)}{144(3^{4K})Ns^2}.
	\end{align}
\end{IEEEproof}

Focusing on the case where the coefficients follow the separable sparsity model, the next theorem provides a lower bound on the minimax risk for coefficients selected according to \eqref{crack} and \eqref{gaussian}.

\begin{theorem} \label{thm_3}
Consider a KS dictionary learning problem with $N$ i.i.d. observations generated according to model~\eqref{eq:model}. Suppose the true dictionary satisfies \eqref{Dclass} for some $r$ and fixed reference dictionary satisfying \eqref{eq:D_0}. If the reference coordinate dictionaries $\{\D_{0,k}, k \in [K]\}$ satisfy $\RIP(s,\frac{1}{2})$ and the random coefficient vector $\x$ is selected according to \eqref{crack} and \eqref{gaussian}, we have the following lower bound on $\eps^*$:
	\begin{align} \label{eq:thm_3}
	\eps^* &\geq \frac{t}{4}
	\min \bigg\{
		p ,
		\frac{r^2}{2K},
		\frac{\sigma^4 p}{36(3^{4K})Ns^2\sigma_a^4} \nonumber \\
	&\qquad \
		\bigg(c_1\bigg(\sum_{k \in [K]} (m_k-1)p_k\bigg)
		-\frac{1}{2}\log_2 2K -2 \bigg) \bigg\},
	\end{align}
for any $0<t<1$ and any $0<c_1<\dfrac{1-t}{8\log 2}$.
\end{theorem}

We state a variation of Lemma~\ref{lemma_Sp_I_UB} necessary for the proof of Theorem~\ref{thm_3}. The proof of the lemma is provided in the appendix.

\begin{lemma}\label{lemma_Sp_II_UB} Consider the generative model in \eqref{eq:model}. Fix $r>0$ and reference dictionary $\D_0$ satisfying \eqref{eq:D_0}. Then, there exists a set of dictionaries $\cD_L\subseteq \cD$ of cardinality $L=2^{\flr{c_1(\sum_{k \in [K]}(m_k-1)p_k)-\frac{K}{2}\log_2(2K)}}$ such that for any $0<t<1$, any $0<c_1<\frac{t^2}{8\log 2}$, any $\eps'>0$ satisfying
	\begin{align} \label{eps:3}
	0<\eps'\leq r^2 \min \left\{1,\frac{r^2}{2Kp}\right\},
	\end{align}
and any $l,l' \in [L]$, with $l \neq l'$, we have
\begin{align} \label{8eps_2}
\frac{2p}{r^2}(1-t)\eps' \leq \|\D_l-\D_{l'}\|_F^2\leq \frac{4Kp}{r^2}\eps'.
\end{align}
Furthermore, assuming the coefficient matrix $\X$ is selected according to \eqref{crack} and \eqref{gaussian}, the reference coordinate dictionaries $\{\D_{0,k}, k \in [K]\}$ satisfy $\RIP(s_k,\frac{1}{2})$, and considering side information $\T(\X)=\supp(\X)$, we have:
	\begin{align} \label{eq:I_3}
	I(\Y;l|\T(\X))&\leq 36( 3^{4K}  ) \left(\frac{\sigma_a}{\sigma}\right)^4 \frac{Ns^2}{r^2}\eps'.
	\end{align}
\end{lemma}

\begin{IEEEproof} [ Proof of Theorem \ref{thm_3}]
The proof of Theorem \ref{thm_3} follows similar steps as the proof of Theorem \ref{thm_2}. The dissimilarity arises in the condition in \eqref{eps:3} for Lemma~\ref{lemma_Sp_II_UB}, which is different from the condition in \eqref{eps:2} for Lemma~\ref{lemma_Sp_I_UB}. This changes the range for the minimax risk $\eps^*$ in which the lower bound in \eqref{thm_2.1_eps} holds.
\end{IEEEproof}

In the next section, we provide a simple KS dictionary learning algorithm for $2$nd-order tensors and study the corresponding dictionary learning MSE.

\section{Partial Converse}
\label{sec:prtl_cnvrse}
In the previous sections, we provided lower bounds on the minimax risk for various coefficient vector distributions and corresponding side information.
We now study a special case of the problem and introduce an algorithm that achieves the lower bound in Corollary \ref{cor_1} (order-wise) for 2nd-order tensors. This demonstrates that our obtained lower bounds are tight in some cases.

\begin{theorem} \label{thm:PrtCnvrs}
Consider a dictionary learning problem with $N$ i.i.d observations according to model~\eqref{eq:model} for $K=2$ and let the true dictionary satisfy \eqref{Dclass} for $\D_0=\I_p$ and some $r>0$. Further, assume the random coefficient vector  $\x$ is selected according to \eqref{swiss}, $\x\in \{-1,0,1\}^p$, where the probabilities of the nonzero entries of $\x$ are arbitrary.
Next, assume noise standard deviation $\sigma$ and express the KS dictionary as
	\begin{align} \label{eq:D_A_B}
	\D = (\I_{p_1}+\Delt_1)\otimes (\I_{p_2}+\Delt_2),
\end{align}		
where $p=p_1p_2$, $\|\Delt_1\|_F \leq r_1$ and $\|\Delt_2\|_F \leq r_2$. Then, if the following inequalities are satisfied:
	\begin{align} \label{eq:PC_c_r}
	r_1\sqrt{p_2}+r_2\sqrt{p_1}+r_1r_2 &\leq r, \nonumber \\
	(r_1+r_2+r_1r_2)\sqrt{s} &\leq 0.1 \nonumber \\
	\max\lr{\frac{r_1^2}{p_2},\frac{r_2^2}{p_1} }& \leq \frac{1}{3N}, \nonumber \\
	\sigma &\leq 0.4,
	\end{align}
there exists a dictionary learning scheme whose MSE satisfies
	\begin{align}
	\bbE_\Y\lr{\|\wh{\D}(\Y)-\D\|_F^2}
	\leq  \frac{8p}{N} & \left( \frac{p_1m_1+p_2m_2}{m\SNR}+3(p_1+p_2) \right) \nonumber \\
		&\ + 8p\exp\left( -\frac{0.08pN}{\sigma^2} \right),
	\end{align}
for any $\D \in \cX(\D_0,r)$ that satisfies~\eqref{eq:D_A_B} .
\end{theorem}

To prove Theorem~\ref{thm:PrtCnvrs}, we first introduce an algorithm to learn a KS dictionary for 2nd-order tensor data. Then, we analyze the performance of the proposed algorithm and obtain an upper bound for the MSE in the proof of Theorem~\ref{thm:PrtCnvrs}, which is provided in the appendix.\footnote{Theorem~\ref{thm:PrtCnvrs} also implicitly uses the assumption that 
$\max\lr{p_1,p_2} \leq N$. } Finally, we provide numerical experiments to validate our obtained results.

\subsection{KS Dictionary Learning Algorithm}
\label{sec:KS_learn_alg}

We analyze a remarkably simple, two-step estimator that begins with thresholding the observations and then ends with estimating the dictionary. Note that unlike traditional dictionary learning methods, our estimator does not perform iterative alternating minimization.

\paragraph{Coefficient Estimate} We utilize a simple thresholding technique for this purpose. For all $n \in [N]$:
	\begin{align} \label{t_alg}
	\wh{\x}_n = (\wh{x}_{n,1},\dots,\wh{x}_{n,p} )^\top, \ \wh{x}_{n,l}
		= \begin{cases}
   		 1   & \quad \text{if } y_{n,l}>0.5,\\
  		 -1  & \quad \text{if } y_{n,l}<-0.5,\\
    	 0  & \quad \text{otherwise}.
  	\end{cases}
	\end{align}
\paragraph{Dictionary Estimate}
Denoting $\A \triangleq \I_{p_1}+\Delt_1$ and $\B \triangleq \I_{p_2}+\Delt_2$, we can write $\D = \A \otimes \B$. We estimate the columns of $\A$ and $\B$ separately.
To learn $\A$, we take advantage of the Kronecker structure of the dictionary and divide each observation $\y_n \in \bbR^{p_1p_2}$ into $p_2$ observations $\y_{(n,j)}'\in \bbR^{p_1}$:
	\begin{align}
	\y_{(n,j)}' = \lr{y_{n,p_2i+j}}_{i=0}^{p_1-1}, \ j\in[p_2], \ n \in [N].
	\end{align}
This increases the number of observations to $N p_2$.
We also divide the original and estimated coefficient vectors:
	\begin{align} \label{eq:xp}
	\x_{(n,j)}' &= \lr{x_{n,p_2i+j}}_{i=0}^{p_1-1},\nonumber \\
	\wh{\x}_{(n,j)}' &= \lr{\wh{x}_{n,p_2i+j}}_{i=0}^{p_1-1},
		\ j\in[p_2], \ n \in [N].
	\end{align}
Similarly, we define new noise vectors:
	\begin{align} \label{eq:no_upd}
	\boldsymbol{\eta}_{(n,j)}' = \lr{\eta_{n,p_2i+j}}_{i=0}^{p_1-1}, \ j\in [p_2], \ n \in [N].
	\end{align}

To motivate the estimation rule for the columns of $\A$,  let us consider the original dictionary learning formulation, $\y_n = \D\x_n + \boldsymbol{\eta}_n$, which we can rewrite as $\y_n = \x_{n,l} \bd_l + \sum_{i \neq l} \x_{n,i} \bd_i +   \boldsymbol{\eta}_n$. Multiplying both sides of the equation by $\x_{n,l}$ and summing up over all training data, we get $\sum_{n=1}^N \x_{n,l} \y_n = \sum_{n=1}^N (\x_{n,l}^2 \bd_l + \sum_{i \neq l} \x_{n,l}\x_{n,i} \bd_i +   \x_{n,l}\boldsymbol{\eta}_n)$. Using the facts $\bbE_\x \big\{\x_{n,l}^2\big\} = \frac{s}{p}$, $\bbE_\x\lr{\x_{n,l}\x_{n,i}} = 0$ for $l \neq i$, and $\bbE_{\x,\boldsymbol{\eta}}\lr{\x_{n,l}\boldsymbol{\eta}_n}=0$, we get the following approximation, $\bd_l \approx \frac{p}{Ns} \sum_{n=1}^N x_{n,l} \y_n$.\footnote{Notice that the i.i.d. assumption on $\x_{n,l}$'s is critical to making this approximation work.} This suggests that for estimating the columns of $\A$, we can utilize the following equation:
	\begin{align} \label{eq:a_update}
	\wt{\ba}_l = \frac{p_1}{Ns}\sum_{n=1}^{N} \sum_{j=1}^{p_2} x_{(k,j),l}'\y_{(n,j)}',\ l \in [p_1].
	\end{align}

To estimate the columns of $\B$, we follow a different procedure to divide the observations. Specifically, we divide each observation $\y_n \in \bbR^{p_1p_2}$ into $p_1$ observations $\y_{(n,j'')}\in \bbR^{p_2}$:
	\begin{align}
	\y_{(n,j)}'' = \lr{y_{n,i+p_1(j-1)}}_{i=1}^{p_2}, \ j \in [p_1], \ n \in [N].
	\end{align}
This increases the number of observations to $Np_1$. The coefficient vectors are also divided similarly:
	\begin{align} \label{eq:xpp}
	\x_{(n,j)}'' &= \lr{x_{k,i+p_1(j-1)}}_{i=0}^{p_1-1}, \nonumber \\
	\wh{\x}_{(n,j)}''& = \lr{\wh{x}_{n,i+p_1(j-1)}}_{i=0}^{p_1-1},
		\ j \in [p_1], \ n \in [N].
	\end{align}
Similarly, we define new noise vectors:
	\begin{align}
	\boldsymbol{\eta}_{(n,j)}'' = \lr{\eta_{n,i+p_1(j-1)}}_{i=1}^{p_2}, \ j\in [p_1], \ n \in [N].
	\end{align}
Finally, using similar heuristics as the estimation rule for columns of $\A$, the estimate for columns of $\B$ can be obtained using the following equation:
	\begin{align} \label{eq:b_update}
	\wt{\bb}_l = \frac{p_2}{Ns}\sum_{n=1}^{N}\sum_{j = 1}^{p_1} x_{(n,j),l}''\y_{(n,j)}'', \ l \in [p_2].
	\end{align}

The final estimate for the recovered dictionary is
	\begin{align}
	\wh{\D} &= \wh{\A}\otimes \wh{\B}, \nonumber \\
	\wh{\A} &= (\wh{\ba}_1,\dots,\wh{\ba}_{p_1}), \quad
		\wh{\ba}_l = P_{\mathcal{B}_1}(\wt{\ba}_l ),\nonumber \\
	\wh{\B} &= (\wh{\bb}_1,\dots,\wh{\bb}_{p_2}), \quad
	    \wh{\bb}_l = P_{\mathcal{B}_1}(\wt{\bb}_l), \label{update_b}
	\end{align}
where the projection on the closed unit ball ensures that $\|\wh{\ba}_l\|_2\leq 1$ and $\|\wh{\bb}_l\|_2\leq 1$. Note that although projection onto the closed unit ball does not ensure the columns of $\wh{\D}$ to have unit norms, our analysis only imposes this condition on the generating dictionary and the reference dictionary, and not on the recovered dictionary.

\begin{remark}
In addition to the heuristics following \eqref{eq:no_upd}, the exact update rules for $\wt{\A}$ and $\wt{\B}$ in \eqref{eq:a_update} and \eqref{eq:b_update} require some additional perturbation analysis. To see this for the case of $\wt{\A}$, notice that \eqref{eq:a_update} follows from writing $\A \otimes \B$ as $\A \otimes (\I_{p_2} +\Delt_2)$, rearranging each $\y_n$ and $(\A \otimes \I_{p_2}) \x_n$ into $\y'_{(n,j)}$'s and $\A \x_{(n,j)}'$'s, and using them to update $\wt{\A}$. In this case, we treat $(\A \otimes \Delt_2)\x_n$ as a perturbation term in our analysis.
A similar perturbation term appears in the case of the update rule for $\wt{\B}$. The analysis for dealing with these perturbation terms is provided in the appendix.
\end{remark}

\subsection{Empirical Comparison to Upper Bound}

\begin{figure}
	\centering
	\begin{subfigure}{0.49\textwidth}
		\includegraphics[width=\textwidth]{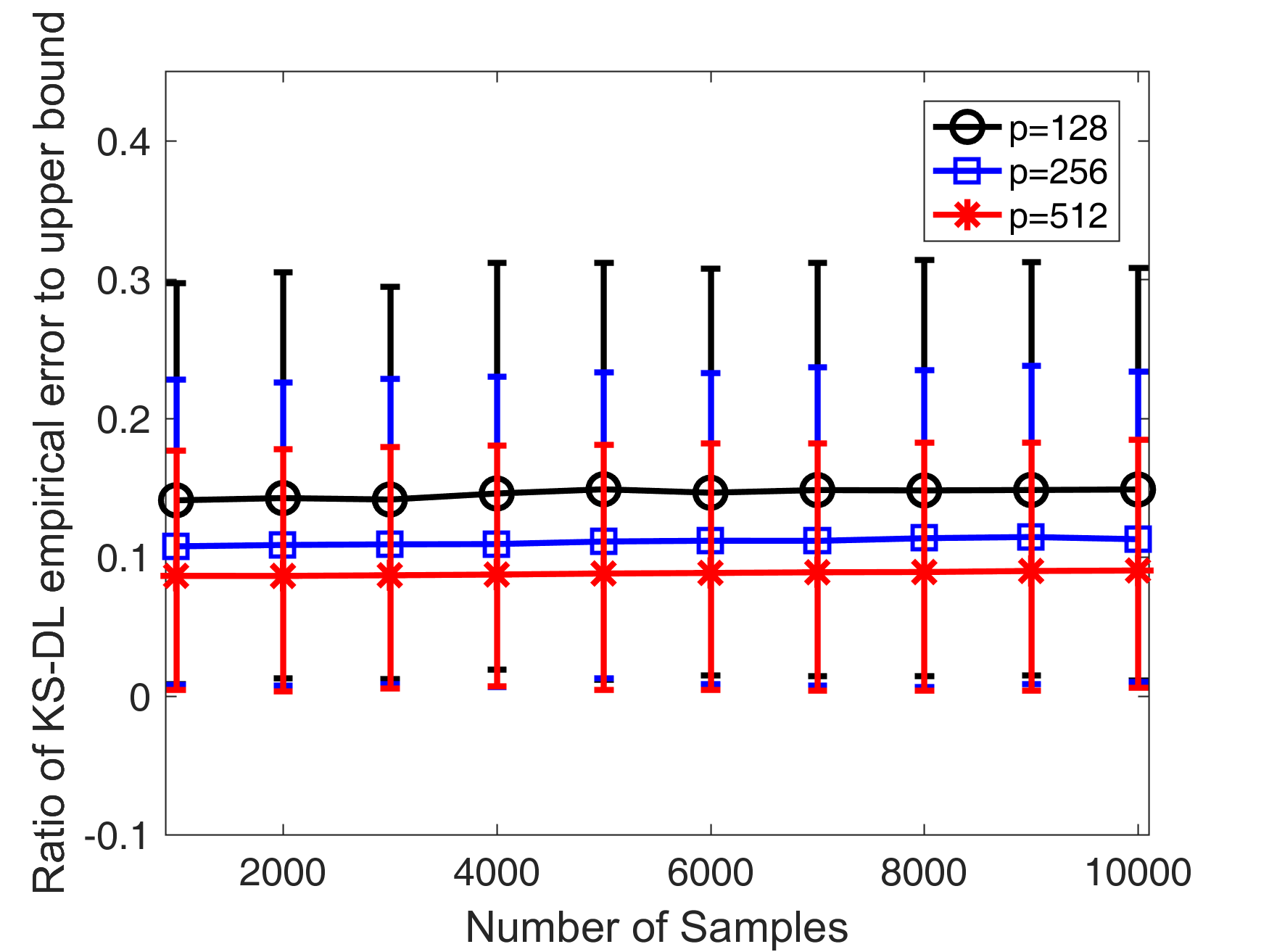}
		\caption{}
		\label{fig:simulation_ratio_Emp_UB}
	\end{subfigure}
	\begin{subfigure}{0.49\textwidth}
		\includegraphics[width=\textwidth]{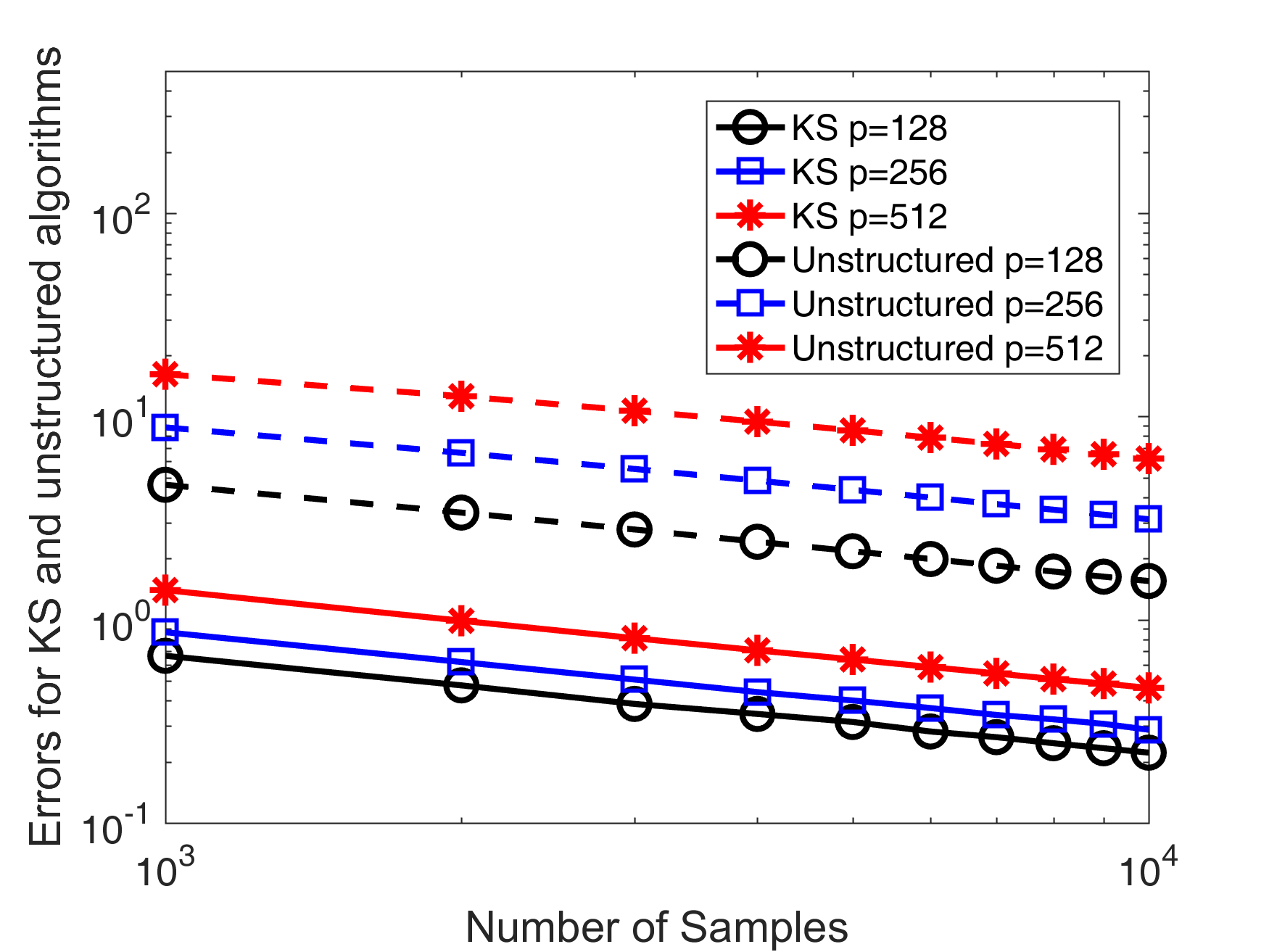}
		\caption{}
		\label{fig:simulation_ratio_Emp_KS_UnS}
	\end{subfigure}
	\caption{Performance summary of KS dictionary learning algorithm for $p=\{128,256,512\}$, $s=5$ and $r=0.1$. (a) plots the ratio of the empirical error of our KS dictionary learning algorithm to the obtained error upper bound along with error bars for generated square KS dictionaries, and (b) shows the performance of our KS dictionary learning algorithm (solid lines) compared to the unstructured learning algorithm proposed in \cite{jung2015minimax} (dashed lines).}
	\label{fig:simulation}
\end{figure}

We are interested in empirically seeing whether our achievable scheme matches the minimax lower bound when learning KS dictionaries. To this end, we implement the preceding estimation algorithm for 2nd-order tensor data.

Figure~\ref{fig:simulation_ratio_Emp_UB} shows the ratio of the empirical error of the proposed KS dictionary learning algorithm in Section~\ref{sec:KS_learn_alg} to the obtained upper bound in Theorem~\ref{thm:PrtCnvrs} for 50 Monte Carlo experiments. This ratio is plotted as a function of the sample size for three choices of the number of columns $p$: $128$, $256$, and $512$. The experiment shows that the ratio is approximately constant as a function of sample size, verifying the theoretical result that the estimator meets the minimax bound in terms of error scaling as a function of sample size. Figure~\ref{fig:simulation_ratio_Emp_KS_UnS} shows the performance of our KS dictionary learning algorithm in relation to the unstructured dictionary learning algorithm provided in~\cite{jung2015minimax}. It is evident that the error of our algorithm is significantly less than that for the unstructured algorithm for all three choices of $p$. This verifies that taking the structure of the data into consideration can indeed lead to lower dictionary identification error.

\section{Discussion}
\label{sec:discussion}

We now discuss some of the implications of our results. Table \ref{table:1} summarizes the lower bounds on the minimax rates from previous papers and this work. The bounds are given in terms of  the number of component dictionaries $K$, the dictionary size parameters ($m_k$'s and $p_k$'s), the coefficient distribution parameters, the number of samples $N$, and $\SNR$, which is defined as
	\begin{align}
	\SNR = \dfrac{\bbE_\x \lr{ \|\x\|_2^2}}{\bbE_{\boldsymbol{\eta}}
		\lr{ \|\boldsymbol{\eta}\|_2^2} }
		 = \dfrac{\tr(\Sig_x)}{m\sigma^2}.
	\end{align}	
These scalings result hold for sufficiently large $p$ and neighborhood radius $r$.

\begin{table*}
\setlength{\extrarowheight}{5pt}
\renewcommand{\arraystretch}{1.3}
\caption{Order-wise lower bounds on the minimax risk for various coefficient distributions}
\label{table:1}
\centering
\begin{tabular}{|l|C{2.4cm}|c|c|}
\hline
\diagbox[width=3.5cm]{\textbf{Distribution}}{\textbf{Dictionary}}
& Side Information $\T(\X)$ & Unstructured~\cite{jung2015minimax}  & Kronecker (this paper) \\
\hline
	1. General  &
	$\X$ &
	$\dfrac{ \sigma^2mp}{N \|\Sig_x\|_2}$ &
	$\dfrac{\sigma^2(\sum_{k \in [K]} m_k p_k)}{NK \|\Sig_x\|_2}$
	\\[2ex]
\hline
	2. Sparse  &
	$\X$ &
	$\dfrac{ p^2}{N \SNR}$ &
	$\dfrac{p(\sum_{k \in [K]} m_k p_k)}{N K m \SNR}$
	\\[2ex]
\hline
	3. Gaussian Sparse &
	$\supp(\X)$ &
	$\dfrac{p^2}{Nm\SNR^2}$ &
	$\dfrac{p(\sum_{k \in [K]} m_k p_k)}{3^{4K}Nm^2\SNR^2}$
	\\[2ex]
\hline
\end{tabular}
\end{table*}

\textbf{Comparison of minimax lower bounds for unstructured and KS dictionary learning:}
Compared to the results for the unstructured dictionary learning problem~\cite{jung2015minimax}, we are able to decrease the lower bound for various coefficient distributions by reducing the scaling $\Omega(mp)$ to $\Omega(\sum_{k \in [K]} m_k p_k)$ for KS dictionaries. This is intuitively pleasing since the minimax lower bound has a linear relationship with the number of degrees of freedom of the KS dictionary, which is $\sum_{k \in [K]} m_k p_k$.

The results also show that the minimax risk decreases with a larger number of samples, $N$, and increased number of tensor order, $K$. By increasing $K$, we are shrinking the size of the class of dictionaries in which the parameter dictionary lies, thereby simplifying the problem.

Looking at the results for the general coefficient model in the first row of Table~\ref{table:1}, the lower bound for any arbitrary zero-mean random coefficient vector distribution with covariance $\Sig_x$ implies an inverse relationship between the minimax risk and $\SNR$ due to the fact that $\|\Sig_x\|_2 \leq \tr(\Sig_x)$.

\textbf{Comparison of general sparse and Gaussian sparse coefficient distributions:}
Proceeding to the sparse coefficient vector model in the second row of Table~\ref{table:1}, by replacing $\Sig_x$ with the expression in \eqref{sig_iid} in the minimax lower bound for the general coefficient distribution, we obtain the second lower bound given in \eqref{eq:cor_1}. Recall that for $s$-sparse coefficient vectors,
	\begin{align}
	\SNR = \dfrac{s\sigma_a^2}{m\sigma^2}.
	\end{align}
Using this definition of $\SNR$ in \eqref{eq:cor_1}, we observe a seemingly counter-intuitive increase in the MSE of order $\Omega\left(p/s\right)$ in the lower bound in comparison to the general coefficient model. However, this increase is due to the fact that we do not require coefficient vectors to have constant energy; because of this, $\SNR$ decreases for $s$-sparse coefficient vectors.

Next, looking at the third row of Table~\ref{table:1}, by restricting the class of sparse coefficient vector distributions to the case where non-zero elements of the coefficient vector follow a Gaussian distribution according to \eqref{gaussian}, we obtain a minimax lower bound that involves less side information than the prior two cases. However, we do make the assumption in this case that reference coordinate dictionaries satisfy $\RIP(s,\frac{1}{2})$. This additional assumption has two implications: (1) it introduces the factor of $1/3^{4K}$ in the minimax lower bound, and (2) it imposes the following condition on the sparsity for the ``random sparsity'' model: $s \leq \min_{k \in [K]} \{p_k\} $. Nonetheless, considering sparse-Gaussian coefficient vectors, we obtain a minimax lower bound that is tighter than the previous bound for some $\SNR$ values. Specifically, in order to compare bounds obtained in \eqref{eq:cor_1} and \eqref{eq:thm_2} for sparse and sparse-Gaussian coefficient vector distributions, we fix $K$. Then in high $\SNR$ regimes, i.e., $\SNR = \Omega( 1/m)$, the lower bound in \eqref{eq:cor_1} is tighter, while \eqref{eq:thm_2} results in a tighter lower bound in low $\SNR$ regimes, i.e., $\SNR = \mc{O}(1/m)$, which correspond to low sparsity settings.

\textbf{Comparison of random and separable sparse coefficient models:}
We now focus on our results for the two sparsity pattern models, namely, random sparsity and separable sparsity, for the case of sparse-Gaussian coefficient vector distribution. These results, which are reported in \eqref{eq:thm_2} and \eqref{eq:thm_3}, are almost identical to each other, except for the first term in the minimization. In order to understand the settings in which the separable sparsity model in \eqref{crack}---which is clearly more restrictive than the random sparsity model in \eqref{swiss}---turns out to be more advantageous, we select the neighborhood radius $r$ to be of order $\mc{O}(\sqrt{p})$; since we are dealing with dictionaries that lie on the surface of a sphere with radius $\sqrt{p}$, this effectively ensures $\cX(\D_0,r) \approx \cD$. In this case, it can be seen from \eqref{eq:thm_2} and \eqref{eq:thm_3} that if $s = \Omega(K)$ then the separable sparsity model gives a better minimax lower bound. On the other hand, the random sparsity model should be considered for the case of $s = \mc{O}(K)$ because of the less restrictive nature of this model.

\textbf{Achievability of our minimax lower bounds for learning KS dictionaries:} To this end, we provided a simple KS dictionary learning algorithm in Section~\ref{sec:prtl_cnvrse} for the special scenario of 2-dimensional tensors and analyzed the corresponding MSE, $\bbE_\Y\big\{ \|\wh{\D}(\Y) - \D\|_F^2\big\}$. In terms of scaling, the upper bound obtained for the MSE in Theorem~\ref{thm:PrtCnvrs} matches the lower bound in Corollary \ref{cor_1} provided $p_1+p_2 < \frac{m_1p_1 + m_2p_2}{m\SNR}$ holds. This result suggests that more general KS dictionary learning algorithms may be developed to achieve the lower bounds reported in this paper.

\section{Conclusion}
\label{sec:conclusion}

In this paper we followed an information-theoretic approach to provide lower bounds for the worst-case mean-squared error (MSE) of Kronecker-structured dictionaries that generate $K$th-order tensor data.
To this end, we constructed a class of Kronecker-structured dictionaries in a local neighborhood of a fixed reference Kronecker-structured dictionary. Our analysis required studying the mutual information between the observation matrix and the dictionaries in the constructed class. To evaluate bounds on the mutual information, we considered various coefficient distributions and interrelated side information on the coefficient vectors and obtained corresponding minimax lower bounds using these models.
In particular, we established that estimating Kronecker-structured dictionaries requires a number of samples that needs to grow only linearly with the sum of the sizes of the component dictionaries ($\sum_{k \in [K]} m_k p_k$), which represents the true degrees of freedom of the problem. We also demonstrated that for a special case of $K = 2$, there exists an estimator whose MSE meets the derived lower bounds. While our analysis is local in the sense that we assume the true dictionary belongs in a local neighborhood with known radius around a fixed reference dictionary, the derived minimax risk effectively becomes independent of this radius for sufficiently large neighborhood radius.

Future directions of this work include designing general algorithms to learn Kronecker-structured dictionaries that achieve the presented lower bounds. In particular, the analysis in \cite{gribonval2015sample} suggests that restricting the class of dictionaries to Kronecker-structured dictionaries may indeed yield a reduction in the sample complexity required for dictionary identification by replacing a factor $mp$ in the general dictionary learning problem with the box counting dimension of the dictionary class~\cite{gribonval2014sparse}.

\section{Acknowledgement}
The authors would like to thank Dr. Dionysios Kalogerias for his helpful comments.

\appendix
\begin{IEEEproof}[Proof of Lemma \ref{lemma_McD}]
Fix $L > 0$ and $\alpha > 0$.
For a pair of matrices $\A_l$ and $\A_{l'}$, with $l \neq l'$, consider the vectorized set of entries $\ba_l=\vect(\A_l)$ and $\ba_{l'}=\vect(\A_{l'})$ and define the function
	\begin{align}
	f(\ba_{l}^\top, \ba_{l'}^\top) &\triangleq \left|\ip{\A_l} {\A_{l'}}\right|
		= \left|\ip{ \ba_{l} }{ \ba_{l'} } \right|.
	\end{align}
For $\wt{\ba} \triangleq (\ba^\top_l,\ba^\top_{l'}) \in \bbR^{2mp}$, write $\wt{\ba} \sim \wt{\ba}'$ if $\wt{\ba}'$ is equal to $\wt{\ba}$ in all entries but one.
Then $f$ satisfies the following bounded difference condition:
	\begin{align}
	\sup_{\wt{\ba}  \sim \wt{\ba}'} \left| f(\wt{\ba} ) - f(\wt{\ba}') \right|
	&=(\alpha-(-\alpha))\alpha =2\alpha^2.
	\end{align}
Hence, according to McDiarmid's inequality~\cite{DubhashiP:09book}, for all $\beta>0$, we have
	\begin{align}
	\bbP\left(\left|\ip{\A_l} {\A_{l'}}\right|\geq \beta \right)
	&\leq
		2\exp\left(\frac{-2\beta^2}{\sum_{i=1}^{2mp} (2\alpha^2)^2} \right)\nonumber \\
	&=
		2\exp\left(-\frac{\beta^2}{4\alpha^4mp}\right).
	\end{align}
Taking a union bound over all pairs $l,l'\in [L], l \neq l'$, we have
	\begin{align} \label{eq:p1}
	\bbP\left(\exists (l,l') \in [L] \times [L], l \neq l': \left|\ip{\A_l} {\A_{l'}}\right| \geq \beta\right) \nonumber \\
	\leq 2L^2\exp\left(-\frac{\beta^2}{4\alpha^4mp} \right) .
	\end{align}
\end{IEEEproof}

\begin{IEEEproof}[Proof of Lemma \ref{lemma_2}]
Fix $r > 0$ and $t \in (0,1)$. Let $\D_0$ be a reference dictionary satisfying \eqref{eq:D_0}, and let $\{\U_{(k,j)}\}_{j=1}^{p_k} \in \bbR^{m_k\times m_k}$, $k\in[K]$, be arbitrary unitary matrices satisfying
	\begin{align}
	\bd_{(k,0),j} &= \U_{(k,j)}\mb{e}_1,
	\end{align}
where $\bd_{(k,0),j}$ denotes the $j$-th column of $\D_{(k,0)}$.

To construct the dictionary class $D_L \subseteq \cX(\D_0, r)$, we follow several steps. We consider sets of
	\begin{align} \label{eq:L_i}
	L_k =  2^{\flr{c_1(m_k-1)p_k - \frac{1}{2}\log_2 2K}}
	\end{align}
generating matrices $\G_{(k,l_k)}$:
\begin{align}
\G_{(k,l_k)}\in \left\{-\frac{1}{r^{1/K}\sqrt{(m_k-1)}},\frac{1}{r^{1/K}\sqrt{(m_k-1)}}\right\}^{(m_k-1)\times p_k}
\end{align}
for $k \in [K]$ and $l_k\in [L_k]$. According to Lemma \ref{lemma_McD}, for all $k\in [K]$ and any $\beta > 0$, the following relation is satisfied:
	\begin{align} \label{eq:p_i}
	\bbP\left(\exists (l_k,l'_k) \in [L_k] \times [L_k], l \neq l': \left|\ip{\G_{(k,l_k)}}{ \G_{(k,l'_k)}}\right| \geq \beta\right) \nonumber \\
	\leq 2L_k^2\exp\left(-\frac{r^{4/K}(m_k-1)\beta^2}{4p_k} \right) .
	\end{align}
To guarantee a simultaneous existence of $K$ sets of generating matrices satisfying
		\begin{align}
	\left|\ip{\G_{(k,l_k)}}{ \G_{(k,l'_k)}}\right| \leq \beta, \quad  k \in [K],
	\end{align}
we take a union bound of \eqref{eq:p_i} over all $k \in [K]$ and choose parameters such that the following upper bound is less than $1$:
	\begin{align}
	 &2KL_k^2\exp\left(-\frac{r^{4/K}(m_k-1)\beta^2}{4p_k} \right)
	 	\nonumber \\
	 &\qquad = \exp\left(-\frac{r^{4/K}(m_k-1)\beta^2}{4p_k}
	 	+2\ln \sqrt{2K}L_k\right),
    \end{align}		
which is satisfied as long as the following inequality holds:
	\begin{align} \label{eq:Li_cond}
	\log_2 L_k < \frac{r^{4/K}(m_k-1)\beta^2}{8p_k\log 2} - \frac{1}{2} - \frac{1}{2} \log_2 K.
	\end{align}
Now, setting $\beta = \dfrac{p_kt}{r^{2/K}}$, the condition in \eqref{eq:Li_cond} holds and there exists a collection of generating matrices that satisfy:
	\begin{align} \label{eq:D1t}
	\left|\ip{\G_{(k,l_k)}}{ \G_{(k,l'_k)}}\right| \leq \frac{p_kt}{r^{2/K}}, \quad  k \in [K],
	\end{align}
for any distinct $l_k,l'_k \in [L_k]$, any $t \in (0,1)$, and any $c_1>0$ such that
	\begin{align} \label{eq:lem1_t}
	c_1 < \frac{t^2}{8 \log 2}.
	\end{align}

We next construct matrices that will be later used for the construction of unit-norm column dictionaries. We construct $\D_{(k,1,l_k)} \in \bbR^{m_k \times p_k}$ column-wise using $\G_{(k,l_k)}$ and unitary matrices $\{\U_{(k,j)}\}_{j=1}^{p_k}$. Let the $j$-th column of $\D_{(k,1,l_k)}$ be given by
	\begin{align} \label{D_2j_D_1j}
	\bd_{(k,1,l_k),j}&=\U_{(k,j)}
	\begin{pmatrix}
		0\\
		\bg_{(k,l_k),j}
	\end{pmatrix}, \quad  k \in [K],
	\end{align}
for any $l_k \in [L_k]$. Moreover, defining
	\begin{align}
	\cD_1 \triangleq \bigg\{ \bigotimes_{k\in [K]} \D_{(k,1,l_k)}: l_k \in [L_k] 	\bigg\},
	\end{align}
and denoting
	\begin{align}
	\cL \triangleq \lr{ (l_1,\dots,l_K): l_k \in [L_k]},
	\end{align}
any element of $\cD_1$ can be expressed as
	\begin{align}\label{eq:D2l}
	\D_{(1,l)} &=  \bigotimes_{k\in [K]} \D_{(k,1,l_k)}, \forall \  l \in [L],
	\end{align}
where $|\cL| = L  \triangleq \prod_{k \in [K]}L_k$ and we associate an $l \in [L]$ with a tuple in $\cL$ via lexicographic indexing. Notice also that
	\begin{align}
	&\norm{\bd_{(1,l),j}}_2^2 \numrel{=}{r_k_l2} \prod_{k \in [K]} 	\
		\norm{\bd_{(k,1,l_k),j}}_2^2
		= \prod_{k \in [K]} \frac{1}{r^{2/K}}
		=\frac{1}{r^2}, \ \text{and} \nonumber\\
	&\norm{\D_{(1,l)}}_F^2 =  \frac{p}{r^2},
	\end{align}
where \eqref{r_k_l2} follows from properties of the Kronecker product. From \eqref{D_2j_D_1j}, it is evident that for all $k\in [K]$, $\bd_{(k,0),j}$ is orthogonal to $\bd_{(k,1,l_k),j}$ and consequently, we have
	\begin{align} \label{eq:orth}
	\ip{\D_{(k,0)}}{\D_{(k,1,l_k)}} = 0, \ k \in [K]
	\end{align}
Also,
	\begin{align}
	 &\ip{ \D_{(k,1,l_k)}}{ \D_{(k,1,l'_k)}}  = \sum_{j=1}^{p_k}
	 \ip{ \bd_{(k,1,l_k),j} }{ \bd_{(k,1,l'_k),j} }  \nonumber\\
	&\qquad \qquad  = \sum_{j=1}^{p_k} \ip{ \U_{(k,j)}
	\begin{pmatrix}
		0\\
		\bg_{(k,l_k),j}
	\end{pmatrix}
	}{
	\U_{(k,j)}
	\begin{pmatrix}
		0\\
		\bg_{(k,l'_k),j}
	\end{pmatrix}
	}  \nonumber
	\\
	& \qquad \qquad \numrel{=}{r_hd} \sum_{j=1}^{p_k} \ip{ \bg_{(k,l_k),j} }{ \bg_{(k,l'_k),j} }  \nonumber\\
	& \qquad \qquad = \ip{\G_{(k,l_k)}}{ \G_{(k,l'_k)}}, \label{sumD2_sumD_1}
	\end{align}
where \eqref{r_hd} follows from the fact that $\{\U_{(k,j)}\}$ are unitary. 	
	
Based on the construction, for all $k \in [K]$, $l_k,l'_k \in [L_k]$, $l_k \neq l'_k$, we have
	\begin{align}
	& \norm{\D_{(1,l)} - \D_{(1,l')}} _F^2 \nonumber \\
	&\qquad = \norm{\D_{(1,l)}}_F^2 + \norm{\D_{(1,l')}}_F^2
		-2\ip{\D_{(1,l)}} {\D_{(1,l')}} \nonumber\\
	&\qquad = \frac{p}{r^2}+\frac{p}{r^2} - 2\prod_{k\in[K]}
		\ip{\D_{(k,1,l_k)}}{ \D_{(k,1,l'_k)}}\nonumber\\
	&\qquad \geq 2\bigg( \frac{p}{r^2} - \prod_{k\in[K]}
		\lra{ \ip{\D_{(k,1,l_k)}}{ \D_{(k,1,l'_k)}} } \bigg) \nonumber\\
	&\qquad \numrel{=}{r_D12} 2\bigg( \frac{p}{r^2} - \prod_{k\in[K]}
		\lra{ \ip{\G_{(k,l_k)}}{ \G_{(k,l'_k)}} } \bigg)\nonumber\\
	&\qquad \numrel{\geq}{r_D22p} 2\bigg( \frac{p}{r^2}
		- \prod_{k\in[K]} \frac{p_k}{r^{2/K}} t \bigg) \nonumber \\
	&\qquad = \frac{2p}{r^2} \left( 1  - t^K \right),\label{eq:D22p}
	\end{align}
where \eqref{r_D12} and \eqref{r_D22p} follow from \eqref{sumD2_sumD_1} and \eqref{eq:D1t}, respectively.

We are now ready to define $\cD_L$. The final dictionary class is defined as
	\begin{align}
	\cD_L \triangleq \bigg\{ \bigotimes_{k\in [K]} \D_{(k,l_k)} : l_k \in [L_k]
		\bigg\}
	\end{align}
and any $\D_l \in \cD_L$ can be written as
	\begin{align} \label{eq:Dl_Def}
	\D_l &= \bigotimes_{k \in [K]} \D_{(k,l_k)},
	\end{align}
where $\D_{(k,l_k)}$ is defined as
	\begin{align} \label{eq:D_{i,l_i}}
	\D_{(k,l_k)} \triangleq \et \D_{(k,0)} + \no \D_{(k,1,l_k)}, \quad k \in [K],
	\end{align}
and
	\begin{align} \label{eq:et_no}
	\et \triangleq \sqrt{1-\dfrac{\eps'}{r^2}}, \quad \no \triangleq \sqrt{\dfrac{r^{2/K} \eps'}{r^2}}
	\end{align}
for any
	\begin{align} \label{eq:eps_cond}
	0 < \eps' < \min \lr{r^2,\frac{r^4}{2Kp}},
	\end{align}
which ensures that $1-\frac{\eps'}{r^2} >0$ and $\D_l \in \cX(\D_0,r)$.
Note that the following relation holds between $\et$ and $\no$:
	\begin{align} \label{eq:et_no_r}
	\et^2 + \frac{\no^2}{r^{2/K}} = 1.
	\end{align}
We can expand \eqref{eq:Dl_Def} to facilitate the forthcoming analysis:
	\begin{align}
	\D_l =  \sum_{\bi \in \{0,1\}^K}
		\et^{K-\| \bi \|_1}
		\no ^{\| \bi \|_1}
		\bigg( \bigotimes_{k \in [K]}  \D_{(k,i_k,l_k)}
		\bigg),
	\end{align}
where $\bi \triangleq \lrp{i_1,i_2,\dots,i_K }$ and $\D_{(k,0,l_k)} \triangleq \D_{(k,0)}$. To show $\cD_L \subseteq \cX(\D_0,r)$, we first show that any $\D_l\in \cD_L$ has unit-norm columns. For any $j \in [p]$ and $j_k \in [p_k], k \in [K]$ (associating $j$ with $(j_1,\dots,j_K)$ via lexicographic indexing), we have
	\begin{align}
	\norm{ \bd_{l,j} }_2^2
	&= \prod_{k \in [K]}\norm{ \bd_{(k,l_k),j_k} }_2^2 \nonumber \\
	&= \prod_{k \in [K]} \bigg(\et^2 \|\bd_{(k,0),j_k}\|_2^2
		+ \no^2 \norm{ \bd_{(k,1,l_k),j_k}}_2^2 \bigg) \nonumber\\
	&= \prod_{k \in [K]}  \bigg(\et^2 + \no^2 \big( \frac{1}{r^{2/K}}\big) \bigg) \nonumber \\
	&\numrel{=}{r_d_0} 1, \label{eq:dj_1}
	\end{align}		
where \eqref{r_d_0} follows from \eqref{eq:et_no_r}. Then, we show that $\norm{ \D_l - \D_0}_F \leq r$:
	\begin{align}
	&\norm{ \D_l - \D_0}_F^2 \nonumber \\
	&= \bigg\| \ \D_0 -   \sum_{\bi \in \{0,1\}^K}
		\et^{K-\| \bi \|_1}
		\no ^{\| \bi \|_1}
		\bigotimes_{k \in [K]}\D_{(k,i_k,l_k)}
		\bigg\|_F^2 \nonumber\\
	&= \bigg\| \big( 1-\et^K \big) \D_0
		-  \sum_{\substack{\bi \in \{0,1\}^K\\ \| \bi \|_1 \neq 0 }}
		\et^{K-\| \bi \|_1}
		\no ^{\| \bi \|_1}
		\bigotimes_{k \in [K]}\D_{(k,i_k,l_k)}   \bigg\|_F^2 \nonumber\\
	& = \left( 1-\et^K \right)^2 \norm{\D_0}_F^2 \nonumber \\
	&\qquad  +\sum_{\substack{\bi \in \{0,1\}^K\\ \| \bi \|_1 \neq 0 }}
		\et^{2(K-\| \bi \|_1)}
		\no ^{2\| \bi \|_1}
		\prod_{k \in [K]}\norm{ \D_{(k,i_k,l_k)}}_F^2 . \label{eq:Dl0}
	\end{align}
We will bound the two terms in \eqref{eq:Dl0} separately.
We know
	\begin{align} \label{eq:binom}
	(1-x^n) = (1-x)(1+x+x^2+\dots+x^{n-1}).
	\end{align}
Hence, we have
	\begin{align}
	\left( 1-\et^K \right)^2 \|\D_0\|_F^2
	& = \left( 1-\et^K \right)^2 p \nonumber\\
	& \numrel{\leq}{r_ghsmt_1} \left( 1-\et^K \right)p\nonumber\\
	& \leq \left( 1-\et^{2K} \right) p\nonumber\\
	& \numrel{=}{r_ghsmt_2} \left( 1-\et^2 \right)\left( 1+\et^2+\dots+ \et^{2(K-1)}\right) p\nonumber\\
	& = \frac{\eps'}{r^2}\left( 1+\et^2+\dots+ \et^{2(K-1)}\right)p\nonumber\\
	& \numrel{\leq}{r_ghsmt_3} \frac{Kp \eps'}{r^2}, \label{eq:ghsmt_1}
	\end{align}
where \eqref{r_ghsmt_1} and \eqref{r_ghsmt_3} follow from the fact that $\et<1$ and \eqref{r_ghsmt_2} follows from \eqref{eq:binom}.

Similarly for the second term in \eqref{eq:Dl0},
	\begin{align}
	 & \prod_{k \in [K]}\norm{\D_{(k,i_k,l_k)}}_F^2  \nonumber \\
	 &\qquad = \bigg( \prod_{\substack{k \in [K] \\ i_{k} = 0}}\|\D_{(k,0)}\|_F^2  \bigg)
		\bigg( \prod_{\substack{k \in [K] \\ i_{k} = 1}}\|\D_{(k,1,l_{k})} \|_F^2 \bigg) \nonumber\\
	 &\qquad =
		\bigg(\prod_{\substack{k \in [K]\\ i_{k} =0}} p_{k} \bigg)
		\bigg(\prod_{\substack{k \in [K]\\ i_{k} =1}} \frac{p_{k}}{r^{2/K}} \bigg)
		\nonumber\\
	&\qquad =
		\bigg(\prod_{k \in [K]} p_{k} \bigg)
		\left( \frac{1}{r^{2/K}}\right)^{\|\bi\|_1}.  \label{eq:D_0_l}
	\end{align}
Replacing values for $\et$ and $\no$ from \eqref{eq:et_no} and using \eqref{eq:D_0_l} and the fact that $\prod_{k \in [K]}p_k=p$, we can further reduce the second term in~\eqref{eq:Dl0} to get
	\begin{align}
	&\sum_{\substack{\bi \in \{0,1\}^K\\ \| \bi \|_1 \neq 0 }}
		\et^{2(K-\| \bi \|_1)}
		\no ^{2\| \bi \|_1}
		\prod_{k \in [K]}\|\D_{(k,i_k,l_k)}\|_F^2 \nonumber \\
	&\qquad = p \sum_{k=0}^{K-1}
		{K \choose k}
		\left( 1 - \frac{\eps'}{r^2}\right)^k
		\left( \frac{\eps'}{r^2}\right)^{K-k} \nonumber \\
	& \qquad = p \left(1 - \left( 1 - \frac{\eps'}{r^2}\right)^K\right) \nonumber\\
	& \qquad \numrel{=}{r_binom} p \left( \frac{\eps'}{r^2} \right)
		\left( 1+\left( 1-\frac{\eps'}{r^2} \right)+\dots+ \left( 1-\frac{\eps'}{r^2} \right)^{K-1}\right)\nonumber\\
	& \qquad \leq \frac{K p \eps'}{r^2}, \label{eq:ghsmt_2}
	\end{align}
where \eqref{r_binom} follows from \eqref{eq:binom}.
Adding \eqref{eq:ghsmt_1} and \eqref{eq:ghsmt_2}, we get
	\begin{align}
	\norm{\D_l - \D_0}_F^2
	&\leq \eps'\left( \frac{2Kp}{r^2} \right) \nonumber\\
	&\numrel{\leq}{eps_con} r^2,
	\end{align}
where \eqref{eps_con} follows from the condition in  \eqref{eq:eps_cond}. Therefore, \eqref{eq:dj_1} and \eqref{eq:ghsmt_2} imply that $\cD_L \subseteq \cX(\D_0,r)$.

We now find lower and upper bounds for the distance between any two distinct elements $\D_l,\D_{l'} \in \cD_L$.

\subsubsection{Lower bounding $\|\D_l-\D_{l'}\|_F^2$}

We define the set $\mc{I}_i \subseteq [K]$ where $|\mc{I}_i| = i, i \in [K]$. Then, given distinct $l_k, l'_k, k \in \mc{I}_i$, we have
	\begin{align}
	\bigg\|\bigotimes_{k \in \mc{I}_i} \D_{(k,1,l_k)} - \bigotimes_{k \in \mc{I}_i} \D_{(k,1,l'_k)} \bigg\|_F^2
	&\numrel{\geq}{r_D2gk} \frac{2\lrp{ 1-t^i }}{r^{2i/K}} \prod_{k \in \mc{I}_i} p_k \nonumber \\
	&\geq \frac{2\lrp{1-t}}{r^{2i/K}} \prod_{k \in \mc{I}_i} p_k, \label{eq:D2gK}
	\end{align}
where \eqref{r_D2gk} follows using arguments similar to those made for \eqref{eq:D22p}.

To obtain a lower bound on $\|\D_l -\D_{l'}\|_F^2$, we emphasize that for distinct $l,l' \in [L]$, it does not necessarily hold that $l_k \neq l_k'$ for all $k \in [K]$.  In fact, it is sufficient for $\D_l \neq \D_{l'}$ that only one $k \in [K]$ satisfies $l_k\neq l'_k$.
Now, assume only $K_1$ out of $K$ coordinate dictionaries are distinct (for the case where all smaller dictionaries are distinct, $K_1 = K$). Without loss of generality, we assume $l_1,\dots,l_{K_1}$ are distinct and $l_{K_1+1},\dots, l_K$ are identical across $\D_l$ and $\D_{l'}$. This is because of the invariance of the Frobenius norm of Kronecker products under permutation, i.e.,
	\begin{align}
	\bigg\| \bigotimes_{k\in [K]} \A_k \bigg\|_F
	&= \prod_{k \in [K]} \norm{\A_k}_F
		=\bigg\| \bigotimes_{k\in [K]} \A_{\pi(k)} \bigg\|_F,
	\end{align}
where $\pi(.)$ denotes a permutation of $[K]$.
We then have
	\begin{align}
	&\norm{\D_l - \D_{l'}}_F^2 \nonumber \\
	&\quad = \bigg\|(\D_{(1,l_1)}\otimes \dots \otimes \D_{(K_1,l_{K_1})} \otimes  \nonumber \\
	&\qquad \qquad \qquad \D_{(K_1+1,l_{K_1+1})} \otimes \dots \otimes \D_{(K,l_K)})  \nonumber \\
	&\qquad - (\D_{(1,l'_1)}\otimes \dots \otimes \D_{(K_1,l'_{K_1})} \otimes \nonumber \\
	&\qquad \qquad \qquad \D_{(K_1+1,l_{K_1+1})} \otimes \dots \otimes \D_{(K,l_K)})\bigg\|_F^2 \nonumber \\
	&\quad \numrel{=}{re_1} \bigg\| \bigg( \bigotimes_{k\in [K_1]}\D_{(k,l_k)}
		- \bigotimes_{k\in [K_1]}\D_{(k,l'_k)} ) \bigg)
		\otimes  \nonumber \\
	&\qquad \qquad \qquad
		\D_{(K_1+1,l_{K_1+1})} \otimes \dots \otimes
		\D_{(K,l_K})) \bigg\|_F^2 \nonumber \\
	&\quad = \bigg\|\bigotimes_{k\in [K_1]}\D_{(k,l_k)}
		- \bigotimes_{k\in [K_1]}\D_{(k,l'_k)} \bigg\|_F^2
		\prod_{k=K_1+1}^K \norm{\D_{(k,l_k)}}_F^2 \nonumber \\
	&\quad = \bigg( \prod_{k=K_1+1}^K p_k \bigg)
		\bigg\| \sum_{\substack{\bi \in \{0,1\}^{K_1}\\ \| \bi \|_1 \neq 0 }}
		\et^{K_1-\| \bi \|_1}
		\no ^{\| \bi \|_1}  \nonumber \\
	&\qquad \qquad \qquad
		\bigg( \bigotimes_{k \in [K_1]} \D_{(k,i_k,l_k)}
		 - \bigotimes_{k \in [K_1]} \D_{(k,i_k,l'_k)} \bigg)  \bigg\|_F^2
		 \nonumber\\
	&\quad \numrel{=}{r_orth}
		\bigg( \sum_{\substack{\bi \in \{0,1\}^{K_1}\\ \| \bi \|_1 \neq 0 }}
		\et^{2(K_1-\| \bi \|_1)}
		\no ^{2\| \bi \|_1}
		\prod_{\substack{k \in [K_1] \\ i_k = 0}}\norm{\D_{(k,0)}}_F^2
		 \nonumber \\
	&\qquad \qquad \qquad \quad
	    \bigg\| \bigotimes_{\substack{k \in [K_1] \\ i_k = 1}} \D_{(k,1,l_k)}
		- \bigotimes_{\substack{k \in [K_1] \\ i_k = 1}} \D_{(k,1,l'_k)}  \bigg\|_F^2  \bigg) \nonumber\\
	&\quad \numrel{\geq}{re_3} \bigg( \prod_{k=K_1+1}^K p_k  \bigg)
		\bigg(  \sum_{\substack{\bi \in \{0,1\}^{K_1}\\ \| \bi \|_1 \neq 0 }}
		\et^{2(K_1-\| \bi \|_1)}
		\no ^{2\| \bi \|_1}  \nonumber \\
	&\quad \qquad \qquad
		\bigg( \prod_{\substack{k \in [K_1] \\ i_k = 0}} p_k \bigg)
		\bigg(\frac{2}{r^{2\|\bi\|_1 /K}} \prod_{\substack{k \in [K_1] \\ i_k = 1}} p_k \bigg)
			\left(1-t\right) \bigg)  \nonumber\\
	&\quad \numrel{=}{re_4} 2p \left( 1- t \right) \sum_{k=0}^{K_1-1}
		{K_1 \choose k}
		\left( 1 - \frac{\eps'}{r^2}\right)^{k}
		\left( \frac{\eps'}{r^2} \right)^{K_1-k} \nonumber\\
	&\quad \numrel{=}{re_5} 2p \left( 1- t \right)
		\left( 1 - \left( 1 - \frac{\eps'}{r^2}\right)^{K_1} \right) \nonumber\\
	&\quad \geq 2 p \left( 1 - t \right)
		\left( 1 - \left( 1 - \frac{\eps'}{r^2}\right) \right)  \nonumber\\
	&\quad = \frac{2 p}{r^2} \left( 1 - t\right) \eps',
	\end{align}
where \eqref{re_1} follows from the distributive property of Kronecker products, \eqref{r_orth} follows the fact that terms in the sum have orthogonal columns (from \eqref{eq:Kron_prod} and \eqref{eq:orth}), \eqref{re_3} follows from \eqref{eq:D2gK}, \eqref{re_4} follows from substituting values for $\et$ and $\no$, and \eqref{re_5} follows from the binomial formula.

\subsubsection{Upper bounding $\|\D_l-\D_{l'}\|_F^2$}
In order to upper bound $\|\D_l-\D_{l'}\|_F^2$, notice that
	\begin{align}
	&\|\D_l-\D_{l'}\|_F^2 \nonumber\\
	&= \sum_{\substack{\bi \in \{0,1\}^{K}\\ \| \bi \|_1 \neq 0 }}
		\et^{2(K-\| \bi \|_1)}
		\no ^{2\| \bi \|_1}   \nonumber \\
	&\qquad \qquad \qquad \bigg\|\bigotimes_{k \in [K]} \D_{(k,i_k,l_k)}
		- \bigotimes_{k \in [K]} \D_{(k,i_k,l'_k)}\bigg\|_F^2 \nonumber \\
	& \numrel{\leq}{r_l_tr}
		\sum_{\substack{\bi \in \{0,1\}^{K}\\ \| \bi \|_1 \neq 0 }}
		\et^{2(K-\| \bi \|_1)}
		\no ^{2\| \bi \|_1}   \nonumber \\
	& \qquad \qquad \quad \bigg( \bigg\|\bigotimes_{k \in [K]}
		 \D_{(k,i_k,l_k)}\bigg\|_F +\bigg\| \bigotimes_{k \in [K]} \D_{(k,i_k,l'_k)}\bigg\|_F \bigg)^2 \nonumber \\
	& =4 \sum_{\substack{\bi \in \{0,1\}^{K}\\ \| \bi \|_1 \neq 0 }}
		\et^{2(K-\| \bi \|_1)}
		\no ^{2\| \bi \|_1}
		\bigg\|\bigotimes_{k \in [K]} \D_{(k,i_k,l_k)}\bigg\|_F^2 \nonumber \\
	& = 4 \sum_{\substack{\bi \in \{0,1\}^{K}\\ \| \bi \|_1 \neq 0 }}
		\et^{2(K-\| \bi \|_1)}
		\no ^{2\| \bi \|_1}   \nonumber \\
	& \qquad \qquad \qquad \prod_{\substack{k \in [K] \\ i_{k} = 0}}\|\D_{(k,0)}\|_F^2
		\prod_{\substack{k \in [K] \\ i_{k} = 1}}\|\D_{(k,1,l_{k})} \|_F^2
		\nonumber \\
	& = 4\sum_{\substack{\bi \in \{0,1\}^{K}\\ \| \bi \|_1 \neq 0 }}
		\et^{2(K-\| \bi \|_1)}
		\no ^{2\| \bi \|_1}
		\bigg( \prod_{\substack{k \in [K] \\ i_{k} = 0}} p_{k} \bigg)
		\bigg(\prod_{\substack{k\in [K]\\i_k=1}} \frac{p_{k}}{r^{2/K}}\bigg)
		\nonumber \\
	&\quad \numrel{=}{r_s_et_no} 4p \sum_{k=0}^{K-1}
		{K \choose k}
		\left( 1 - \frac{\eps'}{r^2}\right)^k
		\left( \frac{\eps'}{r^2}\right)^{K-k}\nonumber\\
	&\quad \numrel{\leq}{r_l_ghsmt_2}\frac{4Kp }{r^2} \eps', \label{eq:up_dllp}
	\end{align}		
where \eqref{r_l_tr} follows from the triangle inequality, \eqref{r_s_et_no} follows from substituting values for $\et$ and $\no$, and \eqref{r_l_ghsmt_2} follows from similar arguments as in \eqref{eq:ghsmt_2}.

\subsubsection{Upper bounding $I(\Y;l|\T(\X))$}
We next obtain an upper bound for $I(\Y;l|\T(\X))$ for the dictionary set $\cD_L$ according to the general coefficient model and side information $\T(\X)=\X$.

Assuming side information $\T(\X)=\X$, conditioned on the coefficients $\x_n$, the observations $\y_n$ follow a multivariate Gaussian distribution with covariance matrix $\sigma^2 \I$ and mean vector $\D\x_n$.
From the convexity of the KL divergence~\cite{cover2012elements}, following similar arguments as in~\cite{wainwright2009information,jung2015minimax}, we have
	\begin{align}
	& I(\Y;l|\T(\X)) = I(\Y;l|\X) \nonumber \\
	&\quad = \frac{1}{L} \sum_{l\in [L]} \bbE_\X\bigg\{D_{KL}\bigg(f_{\D_l}(\Y|\X) \big\|
		\frac{1}{L}\sum_{l' \in [L]}f_{\D_{l'}}(\Y|\X)\bigg)\bigg\}\nonumber \\
	&\quad \leq \frac{1}{L^2} \sum_{l,l' \in [L]} \bbE_\X\bigg\{D_{KL}\bigg(f_{\D_l}(\Y|\X) \big\|
		f_{\D_{l'}} (\Y|\X) \bigg)\bigg\}, \label{eq:KL-case1}
	\end{align}
where $f_{\D_l}(\Y|\X)$ is the probability distribution of the observations $\Y$, given the coefficient matrix $\X$ and the dictionary $\D_l$. From Durrieu et al.~\cite{durrieu2012lower}, we have
	\begin{align}
	&D_{KL}\bigg(f_{\D_l}(\Y|\X) \big\| f_{\D_{l'}}(\Y|\X)  \bigg) \nonumber \\
	&\qquad = \sum_{n\in [N]}\frac{1}{2\sigma^2} \norm{(\D_l-\D_{l'})\x_n}^2_2 \nonumber \\
	&\qquad = \sum_{n\in [N]} \frac{1}{2\sigma^2} \tr \lr{(\D_l-\D_{l'})^\top (\D_l-\D_{l'}) \x_n\x_n^\top }. \label{tr}
	\end{align}
Substituting \eqref{tr} in \eqref{eq:KL-case1} results in
	\begin{align}
	&I(\Y;l|\T(\X)) \nonumber \\
	&\quad \leq  \bbE_\X \bigg\{\sum_{n\in [N]} \frac{1}{2\sigma^2}
		\tr \lr{(\D_l-\D_{l'})^\top (\D_l-\D_{l'}) \x_n\x_n^\top } \bigg\} 	
		\nonumber \\
	&\quad = \sum_{n\in [N]} \frac{1}{2\sigma^2}
		\tr \lr{(\D_l-\D_{l'})^\top (\D_l-\D_{l'}) \Sig_x} \nonumber \\
	&\quad \numrel{\leq}{rel9} \sum_{n\in [N]} \frac{1}{2\sigma^2}
		\|\Sig_x\|_2 \|\D_l-\D_{l'}\|_F^2   \nonumber \\
	&\quad \numrel{\leq}{rel10} \frac{N}{2\sigma^2} \|\Sig_x\|_2
		\left( \frac{4Kp\eps'}{r^2} \right) \nonumber \\
	&\quad = \frac{2NKp\|\Sig_x\|_2}{r^2\sigma^2}\eps', \label{UB}
	\end{align}
where \eqref{rel10} follows from \eqref{eq:up_dllp}. To show \eqref{rel9}, we use the fact that for any $\A \in \bbR^{p\times p}$ and $\Sig_x$ with ordered singular values $\sigma_i(\A)$ and $\sigma_i(\Sig_x), i \in [p]$, we have
	\begin{align}
	\tr \lr{\A\Sig_x }
	&\leq \lra{ \tr\lr{\A\Sig_x } }	\nonumber \\
	&\numrel{\leq}{r_tr_neq} \sum_{i=1}^{p} \sigma_i(\A) \sigma_i(\Sig_x) \nonumber \\
	&\numrel{\leq}{r_sig_psd} \sigma_1(\Sig_x)
		\sum_{i=1}^{p} \sigma_i(\A)		\nonumber\\
	&=\|\Sig_x\|_2 \tr\{\A\},
	\end{align}
where \eqref{r_tr_neq} follows from Von Neumann's trace inequality \cite{neumann1937some} and \eqref{r_sig_psd} follows from the positivity of the singular values of $\Sig_x$. The inequality in \eqref{rel9} follows from replacing $\A$ with $(\D_l-\D_{l'})^\top (\D_l-\D_{l'}) $ and using the fact that $\tr\{(\D_l-\D_{l'})^\top (\D_l-\D_{l'}) \}=\|\D_l - \D_{l'}\|_F^2$.
\end{IEEEproof}

\begin{IEEEproof}[Proof of Lemma~\ref{lemma_Sp_I_UB}]
The dictionary class $\cD_L$ constructed in Lemma \ref{lemma_2} is again considered here. Note that \eqref{eps:2} implies $\eps'<r^2$, since $s\geq 1$. The first part of Lemma~\ref{lemma_Sp_I_UB}, up to \eqref{eq:8eps_Sprs}, thus trivially follows from Lemma~\ref{lemma_2}.
In order to prove the second part, notice that in this case the coefficient vector is assumed to be sparse according to \eqref{swiss}. Denoting $\x_{\cS_n}$ as the elements of $\x_n$ with indices $\cS_n\triangleq \supp(\x_n)$, we have observations $\y_n$ as
	\begin{align} \label{eq:y_k_case2}
	\y_n = \D_{l,\cS_n}\x_{\cS_n} + \boldsymbol{\eta}_n.
	\end{align}
Hence conditioned on $\cS_n=\supp(\x_n)$, observations $\y_n$'s are zero-mean independent multivariate Gaussian random vectors with covariances
	\begin{align} \label{eq:cov_case2}
	\Sig_{(n,l)} = \sigma_a^2 \D_{l,\cS_n}\D_{l,\cS_n}^\top +\sigma^2\I_s.
	\end{align}
The conditional MI $I(\Y;l|\T(\X)=\supp(\X))$ has the following upper bound~\cite{wang2010information,jung2015minimax}:
	\begin{align} \label{eq:KL_case2}
	I(\Y;l & |\T(\X)) \leq \bbE_{\T(\X)}
		\bigg\{ \sum_{\substack{n \in [N]\\ l,l' \in [L]}}  \frac{1}{L^2} \nonumber \\
	&\ \qquad \tr \big\{\big[ \Sig_{(n,l)}^{-1} 	- \Sig_{(n,l')}^{-1}\big]
		\big[\Sig_{(n,l)} - \Sig_{(n,l')}\big] \big\}\bigg\} \nonumber\\
	&\leq \rnk \lr{\Sig_{(n,l)} - \Sig_{(n,l')}}  \bbE_{\T(\X)}
		\bigg\{ \sum_{n\in[N]} \frac{1}{L^2}\nonumber \\
	&\qquad  \sum_{l,l' \in [L]}\norm{\Sig_{(n,l)}^{-1}
		- \Sig_{(n,l')}^{-1}}_2 \norm{\Sig_{(n,l)} - \Sig_{(n,l')}}_2 \bigg\}.
	\end{align}
Since $\rnk(\Sig_{(n,l)})\leq s$, $\rnk\{\Sig_{(n,l)} - \Sig_{(n,l')}\} \leq 2s$~\cite{jung2015minimax}.

Next, note that since non-zero elements of the coefficient vector are selected according to \eqref{swiss} and \eqref{gaussian},
we can write the subdictionary $\D_{l,\cS_n}$ in terms of the Khatri-Rao product of matrices:
	\begin{align}
	\D_{l,\cS_n} =  \bigAst_{k \in [K]}  \D_{(k,l_k),\cS_{n_k}},
	\end{align}	
where $\cS_{n_k}=\{j_{n_k}\}_{n_k=1}^s, j_{n_k} \in [p_k]$, for any $k \in [K] $, denotes the support of $\x_n$ according to the coordinate dictionary $\D_{(k,l_k)}$ and $\cS_n$ corresponds to the indexing of the elements of $(\cS_1 \times \dots \cS_K)$. Note that $\D_{l,\cS_n} \in \bbR^{(\prod_{k\in[K]}m_k) \times s}$ and in this case, the $\cS_{n_k}$'s can be multisets.\footnote{Due to  the fact that $\cS_{n_k}$'s can be multisets,  $\D_{(k,l_k),\cS_{n_k}}$'s can have duplicated columns.} We can now write
	\begin{align}
	&\Sig_{(n,l)}  = \nonumber \\
	& \sigma_a^2
		\bigg( \bigAst_{k_1 \in [K]} \D_{(k_1,l_{k_1}),\cS_{n_{k_1}}} \bigg)
		\bigg(\bigAst_{k_2 \in [K]} \D_{(k_2,l_{k_2}),\cS_{n_{k_2}}}  \bigg)^\top
		+ \sigma^2 \I_s.
	\end{align}

We next write
	\begin{align} \label{eq:sigl-siglp}
	&\frac{1}{\sigma_a^2}(\Sig_{(n,l)}-\Sig_{(n,l')} ) \nonumber \\
	&= \bigg( \bigAst_{k_1 \in [K]} \D_{(k_1,l_{k_1}),\cS_{n_{k_1}}} \bigg)
		\bigg(\bigAst_{k_2 \in [K]} \D_{(k_2,l_{k_2}),\cS_{n_{k_2}}}  \bigg)^\top
		\nonumber \\
	&\qquad  - \bigg(\bigAst_{k_1 \in [K]} \D_{(k_1,l'_{k_1}),\cS_{n_{k_1}}}  \bigg)
		 \bigg(\bigAst_{k_2 \in [K]} \D_{(k_2,l'_{k_2}),\cS_{n_{k_2}}} \bigg)^\top    \nonumber \\
    & = \bigg( \sum_{\bi \in \{0,1\}^K}
		\et^{K-\| \bi \|_1}
		\no ^{\| \bi \|_1}
		\bigAst_{k_1 \in [K]}  \D_{(k_1,i_{k_1},l_{k_1}),\cS_{n_{k_1}}}
		\bigg)\nonumber \\
	&\qquad \quad \bigg(\sum_{\bi'\in \{0,1\}^K}
		\et^{K-\| \bi' \|_1}
		\no ^{\| \bi' \|_1}
		\bigAst_{k_2 \in [K]}  \D_{(k_2,i'_{k_2},l_{k_2}),\cS_{n_{k_2}}}
		\bigg)^\top \nonumber \\
    & - \bigg( \sum_{\bi \in \{0,1\}^K}
		\et^{K-\| \bi \|_1}
		\no ^{\| \bi \|_1}
		\bigAst_{k_1 \in [K]}  \D_{(k_1,i_{k_1},l'_{k_1}),\cS_{n_{k_1}}}
		\bigg)\nonumber \\
	&\qquad \quad \bigg(\sum_{\bi' \in \{0,1\}^K}
		\et^{K-\| \bi'\|_1}
		\no ^{\| \bi'\|_1}
		\bigAst_{k_2 \in [K]}  \D_{(k_2,i'_{k_2},l'_{k_2}),\cS_{n_{k_2}}}
		\bigg)^\top \nonumber \\
	&= \sum_{\substack{\bi,\bi' \in \{0,1\}^K \\ \|\bi\|_1+ \|\bi'\|_1 \neq 0}}
		\et^{2K-\|\bi\|_1- \|\bi'\|_1}
		\no^{\|\bi\|_1+ \|\bi'\|_1} \nonumber \\
	& \quad
		\bigg( \bigAst_{k_1 \in [K]}  \D_{(k_1,i_{k_1},l_{k_1}),\cS_{n_{k_1}}}	\bigg)
		\bigg( \bigAst_{k_2 \in [K]}  \D_{(k_2,i'_{k_2},l_{k_2}),\cS_{n_{k_2}}}  \bigg)^\top\nonumber \\
	& \quad -  \sum_{\substack{\bi,\bi' \in \{0,1\}^K \\ \|\bi\|_1+ \|\bi'\|_1 \neq 0}}
		\et^{2K-\|\bi\|_1- \|\bi'\|_1}
		\no^{\|\bi\|_1+ \|\bi'\|_1} \nonumber \\
	& \quad
		\bigg( \bigAst_{k_1\in [K]}  \D_{(k_1,i_{k_1},l'_{k_1}),\cS_{n_{k_1}}}	\bigg)
		\bigg( \bigAst_{k_2\in [K]}  \D_{(k_2,i'_{k_2},l'_{k_2}),\cS_{n_{k_2}}}  \bigg)^\top.
	\end{align}
We now note that
	\begin{align} \label{eq:spctr_n_ast}
	\|\A_1 \ast \A_2 \|_2 & = \|(\A_1 \otimes \A_2)\mb{P}\|_2\nonumber \\
	&\leq \|(\A_1 \otimes \A_2)\|_2\|\mb{P}\|_2\nonumber \\
	&\numrel{=}{r_J} \|\A_1\|_2 \| \A_2\|_2 ,
	\end{align}
where $\mb{P} \in \bbR^{p \times s}$ is a selection matrix that selects $s$ columns of $\A_1 \otimes \A_2$ and $\mb{p}_j = \mb{e}_i$ for $j \in [s], i \in [p]$. Here, \eqref{r_J} follows from the fact that $\|\mb{P}\|_2=1$ ($\mb{P}^\top\mb{P}=\I_s$).
From \eqref{eps:2}, it is apparent that $\sqrt{\dfrac{s\eps'}{r^2}}\leq 1$.
Furthermore,
	\begin{align}
	&\norm{ \D_{(k,0),\cS_{{n_k}}}}_2 \leq \sqrt{\frac{3}{2}}  ,
	\norm{ \D_{(k,1,l_k),\cS_{n_k}}}_2 \leq \sqrt{\frac{s}{r^{2/K}}}, \ k \in [K],\label{A2-B2}
	\end{align}
where the fist inequality in \eqref{A2-B2} follows from the $\RIP$ condition for $\lr{\D_{(0,k)}, k \in [K]}$ and the second inequality follows from the fact that $\| \A \|_2 \leq \| \A\|_F$.
We therefore have
	\begin{align}
	&\frac{1}{\sigma_a^2}  \norm{ \Sig_{(n,l)}-\Sig_{(n,l')} }_2 \nonumber \\
	&\numrel{\leq}{r_tri} 2 \sum_{\substack{\bi,\bi' \in \{0,1\}^K \\ \|\bi\|_1+ \|\bi'\|_1 \neq 0}}
		\et^{2K-\|\bi\|_1- \|\bi'\|_1}
		\no^{\|\bi\|_1+ \|\bi'\|_1} \nonumber \\
	&\qquad
		\bigg\| \bigAst_{k_1 \in [K]}  \D_{(k_1,i_{k_1},l_{k_1}),\cS_{n_{k_1}}}	
		\bigg\|_2
		\bigg\| \bigAst_{k_2 \in [K]}  \D_{(k_2,i'_{k_2},l_{k_2}),\cS_{n_{k_2}}}
		\bigg\|_2
		\nonumber \\
	&\numrel{\leq}{r_DS_D0D1} 2
		\sum_{\substack{\bi \in \{0,1\}^K \\ \|\bi\|_1 \neq 0}}
		\et^{K-\|\bi\|_1}
		\no^{\|\bi\|_1} \nonumber \\
	&\qquad \qquad \prod_{\substack{k_1 \in [K]\\ i_{k_1} = 0}}
		\big\| \D_{(k_1,0),\cS_{n_{k_1}}}	\big\|_2
		\prod_{\substack{k_1 \in [K]\\ i_{k_1} = 1}}
		\big\| \D_{(k_1,1,l_{k_1}),\cS_{n_{k_1}}}	\big\|_2 \nonumber \\
	&\qquad \bigg( \sum_{\bi' \in \{0,1\}^K }
		\et^{K-\|\bi'\|_1}
		\no^{\|\bi'\|_1} \nonumber \\
	&\qquad \qquad  \prod_{\substack{k_2 \in [K]\\ i'_{k_2} = 0}}
		\big\| \D_{(k_2,0),\cS_{n_{k_2}}}	\big\|_2
		\prod_{\substack{k_2 \in [K]\\ i'_{k_2} = 1}}
		\big\| \D_{(k_2,1,l_{k_2}),\cS_{n_{k_2}}}	\big\|_2
		\bigg) \nonumber \\
	&+2 \sum_{\substack{\bi \in \{0,1\}^K \\ \|\bi\|_1 \neq 0}}
		\et^{K-\|\bi\|_1}
		\no^{\|\bi\|_1}  \nonumber \\
	&\qquad \qquad \prod_{\substack{k_1 \in [K]\\ i_{k_1} = 0}}
		\big\| \D_{(k_1,0),\cS_{n_{k_1}}}	\big\|_2
		\prod_{\substack{k_1 \in [K]\\ i_{k_1} = 1}}
		\big\| \D_{(k_1,1,l_{k_1}),\cS_{n_{k_1}}}	\big\|_2 \nonumber \\
	&\bigg( \sum_{\substack{\bi' \in \{0,1\}^K \\ \|\bi'\|_1 \neq 0}}
		\et^{K-\|\bi'\|_1}
		\no^{\|\bi'\|_1}  \nonumber \\
	&\qquad \qquad \prod_{\substack{k_2 \in [K]\\ i'_{k_2} = 0}}
		\big\| \D_{(k_2,0),\cS_{n_{k_2}}}	\big\|_2
		\prod_{\substack{k_2 \in [K]\\ i'_{k_2} = 1}}
		\big\| \D_{(k_2,1,l_{k_2}),\cS_{n_{k_2}}}	\big\|_2
		\bigg) \nonumber \\
	& \numrel{=}{r_e_1} 2\bigg( \sum_{k_1=0}^{K-1}
		{K \choose k_1}
		\et^{k_1}
		\no^{K-k_1}
		\bigg(\sqrt{\frac{3}{2}}\bigg)^{k_1}
		\bigg(\sqrt{\frac{s}{r^{2/K}}}\bigg)^{K-k_1}
		\bigg) \nonumber \\
		&\qquad \quad
	    \bigg( \sum_{k_2=0}^{K}
		{K \choose k_2}
		\et^{k_2}
		\no^{K-k_2}
		\left( \sqrt{\frac{3}{2}}\right)^{k_2}
		\left( \sqrt{\frac{s}{r^{2/K}}}\right)^{K-k_2}
		\bigg) \nonumber \\
	&+ 2 \bigg( \et \sqrt{\frac{3}{2}}\bigg)^K
		\bigg( \sum_{k_2=0}^{K-1}
		{K \choose k_2}
		\et^{k_2}
		\no ^{K-k_2}
		\bigg( \sqrt{\frac{3}{2}}\bigg)^{k_2} \nonumber \\
	&\qquad \qquad \qquad \qquad \qquad \qquad \qquad \qquad \quad
		\bigg( \sqrt{\frac{s}{r^{2/K}}}\bigg)^{K-k_2}	\bigg) \nonumber \\
	& \numrel{\leq}{r_et_no} 2\bigg( \sum_{k_1=0}^{K-1}
		{K \choose k_1}
		\bigg(\sqrt{ \frac{3}{2}} \bigg)^{k_1}
		\bigg( \sqrt{\frac{s\eps'}{r^2}}\bigg)^{K-k_1}
		\bigg) \nonumber \\
		&\qquad \qquad \bigg( \sum_{k_2=0}^{K}
		{K \choose k_2}
		\bigg(\sqrt{\frac{3}{2}} \bigg)^{k_2}
		\bigg( \sqrt{\frac{s\eps'}{r^2} }\bigg)^{K-k_2}
		\bigg) \nonumber \\
	& \quad + 2	\bigg(  \sqrt{\frac{3}{2}} \bigg)^K
		\bigg( \sum_{k_2=0}^{K-1}
		{K \choose k_2}
		\bigg(  \sqrt{\frac{3}{2}}  \bigg)^{k_2}
		\bigg( \sqrt{ \frac{s\eps'}{r^2} } \bigg)^{K-k_2} \bigg) \nonumber \\
	& = 2 \sqrt{ \frac{s\eps'}{r^2}} \bigg(
		\sum_{k_1=0}^{K-1}
		{K \choose k_1}
		\bigg( \sqrt{ \frac{3}{2}}\bigg)^{k_1}
		\bigg( \sqrt{ \frac{s\eps'}{r^2} } \bigg)^{K-1-k_1}
		\bigg) \nonumber \\
	&\qquad \bigg( \sum_{k_2=0}^{K}
		{K \choose k_2}
		\bigg( \sqrt{\frac{3}{2}}\bigg)^{k_2}
		\bigg(\sqrt{ \frac{s\eps'}{r^2} } \bigg)^{K-k_2}
		+\bigg( \sqrt{\frac{3}{2}}\bigg)^K
		\bigg)   \nonumber \\
	& \numrel{\leq}{r_et_no2}  2 \sqrt{ \frac{s\eps'}{r^2}}
		\bigg( \bigg( \sqrt{\frac{3}{2}} \bigg)^{K-1}
		\sum_{k_1=0}^K
		{K \choose k_1} \bigg) \nonumber \\
	&\qquad \qquad \bigg( \bigg(  \sqrt{\frac{3}{2}}+ 1
		\bigg)^K
		+ \bigg(\sqrt{\frac{3}{2} }\bigg)^{K} \bigg)
	    \nonumber \\
	& \leq  2  \sqrt{ \frac{s\eps'}{r^2}}
		\bigg( \bigg( \sqrt{\frac{3}{2}}\bigg)^{K-1} 2^K \bigg)
		\bigg( \bigg( \frac{3}{2}\bigg)^{K} 2^K + \bigg( \frac{3}{2}\bigg)^{K}  \bigg)
	    \nonumber \\	
	& \leq 3^{2K+1} \sqrt{ \frac{s\eps'}{r^2}} ,
		\label{eq:sigma_diff}
	\end{align}
where \eqref{r_tri} follows from triangle inequality, \eqref{r_DS_D0D1} follows from \eqref{eq:spctr_n_ast}, \eqref{r_e_1} follows from \eqref{A2-B2}, \eqref{r_et_no} and \eqref{r_et_no2} follow from replacing the value for $\no$ and the fact that $\et<1$ and $s\eps'/r^2<1$ (by assumption).
Denoting the smallest eigenvalue of $\Sig_{(n,l)}$ as $\lambda_{\min}(\Sig_{(n,l)})$, $\lambda_{\min}(\Sig_{(n,l)})\geq \sigma^2$ holds; thus, we have $\|\Sig_{(n,l)}^{-1}\|_2\leq\frac{1}{\sigma^2}$ and from \cite{golub2012matrix}, we get
	\begin{align}
	\norm{\Sig_{(n,l)}^{-1}-\Sig_{(n,l')} ^{-1}}_2
	&\leq 2\norm{\Sig_{(n,l)}^{-1}}^2_2 \norm{\Sig_{(n,l)}-\Sig_{(n,l')} }_2 \nonumber \\
	&\leq \frac{2}{\sigma^4} \norm{\Sig_{(n,l)}-\Sig_{(n,l')} }_2.
	\end{align}
Now \eqref{eq:KL_case2} can be stated as
	\begin{align}
	I(\Y;l|\T(\X))
	&\leq \frac{4Ns}{\sigma^4 L^2}\sum_{l,l'}\norm{\Sig_{(n,l)}-\Sig_{(n,l')}}_2^2 \nonumber \\
	&\leq \frac{4Ns}{\sigma^4 }\norm{\Sig_{(n,l)}-\Sig_{(n,l')} }_2^2 \nonumber \\
	&\numrel{\leq}{rel11} \frac{4N s}{\sigma^4}
		( 3^{4K+2})  \left(\sigma_a^2\sqrt{ \frac{s\eps'}{r^2}}\right)^2 \nonumber \\
	&= 36( 3^{4K}  )\left(\frac{\sigma_a}{\sigma}\right)^4  \frac{N s^2}{r^2}  \eps', \label{eq:MI_U_S}
	\end{align}
where \eqref{rel11} follow from \eqref{eq:sigma_diff}. Thus, the proof is complete.
\end{IEEEproof}

\begin{IEEEproof}[Proof of Lemma~\ref{lemma_Sp_II_UB}]
Similar to Lemma~\ref{lemma_Sp_I_UB}, the first part of this Lemma trivially follows from Lemma \ref{lemma_2}. Also, in this case the coefficient vector is assumed to be sparse according to \eqref{crack}. Hence, conditioned on $\cS_n=\supp(\x_n)$, observations $\y_n$'s are zero-mean independent multivariate Gaussian random vectors with covariances given by \eqref{eq:cov_case2}. Similar to Lemma~\ref{lemma_Sp_I_UB}, therefore, the conditional MI has the upper bound given in \eqref{eq:KL_case2}. We now simplify this upper bound further.

When non-zero elements of the coefficient vector are selected according to \eqref{crack} and \eqref{gaussian},
we can write the dictionary $\D_{l,\cS_n}$ in terms of the Kronecker product of matrices:
	\begin{align}
	\D_{l,\cS_n} =  \bigotimes_{k \in [K]}  \D_{(k,l_k),\cS_{n_k}} ,
	\end{align}	
where $\cS_{n_k}=\{j_{n_k}\}_{n_k=1}^{s_k}, j_{n_k} \in [p_k]$, for all $k \in [K] $, denotes the support of $\x_n$ on coordinate dictionary $\D_{(k,l_k)}$ and $\cS_n$ corresponds to indexing of the elements of $(\cS_1 \times \dots \times \cS_K)$. Note that $\D_{l,\cS_{n}} \in \bbR^{(\prod_{k\in[K]}m_k) \times s}$. In contrast to coefficient model \eqref{swiss}, in this model the $\cS_{n_k}$'s are not multisets anymore since for each $\D_{(k,l_k)}, k \in [K]$, we select $s_k$ columns at random and $ \D_{(k,l_k),\cS_{n_k}} $ are submatrices of $\D_{(k,l_k)}$. Therefore, \eqref{eq:cov_case2} can be written as
	\begin{align}
	&\Sig_{(n,l)}  = \sigma_a^2
		\bigg( \bigotimes_{k_1 \in [K]} \D_{(k_1,l_{k_1}),\cS_{n_{k_1}}} \bigg) \nonumber \\
	& \qquad \qquad \qquad \qquad
		\bigg( \bigotimes_{k_2 \in [K]}  \D_{(k_2,l_{k_2}),\cS_{n_{k_2}}}\bigg)^\top
		+ \sigma^2 \I_s.
	\end{align}
In order to find an upper bound for $\|\Sig_{(n,l)}-\Sig_{(n,l')}\|_2$, notice that the expression for $\Sig_{(n,l)}-\Sig_{(n,l')}$ is similar to that of \eqref{eq:sigl-siglp}, where $\bigAst$ is replaced by $\bigotimes$. Using the property of Kronecker product that $\|\A_1 \otimes \A_2 \|_2 =\|\A_1 \|_2 \| \A_2 \|_2 $ and the fact that
	\begin{align}
	\norm{ \D_{(k,0),\cS_{{n_k}}} }_2 \leq \sqrt{\frac{3}{2}}  ,
	\norm{ \D_{(k,1,l_k),\cS_{n_k}}}_2 \leq \sqrt{\frac{s_k}{r^{2/K}}} ,
	\forall k \in [K], \label{A2-B2_2}
	\end{align}
we have
	\begin{align}
	&\frac{1}{\sigma_a^2} \norm{ \Sig_{(n,l)}-\Sig_{(n,l')} }_2 \nonumber \\
	& \leq 2 \sum_{\substack{\bi,\bi' \in \{0,1\}^K \\ \|\bi\|_1+ \|\bi'\|_1 \neq 0}}
		\et^{2K-\|\bi\|_1- \|\bi'\|_1}
		\no^{\|\bi\|_1+ \|\bi'\|_1} \nonumber \\
	&\qquad
		\bigg\| \bigotimes_{k_1 \in [K]}  \D_{(k_1,i_{k_1},l_{k_1}),\cS_{n_{k_1}}}	
		\bigg\|_2
		\bigg\| \bigotimes_{k_2 \in [K]}  \D_{(k_2,i'_{k_2},l_{k_2}),\cS_{n_{k_2}}}
		\bigg\|_2 \nonumber \\
	&  = 2\sum_{\substack{\bi \in \{0,1\}^K \\ \|\bi\|_1 \neq 0}}
		\et^{K-\|\bi\|_1}
		\no^{\|\bi\|_1}  \nonumber \\
	&\quad \qquad \prod_{\substack{k_1 \in [K]\\ i_{k_1} = 0}}   \big\| \D_{(k_1,0),\cS_{n_{k_1}}}	
		\big\|_2
		\prod_{\substack{k_1 \in [K]\\ i_{k_1} = 1}}   \big\| \D_{(k_1,1,l_{k_1}),\cS_{n_{k_1}}}	\big\|_2 \nonumber \\
	&\qquad \bigg( \sum_{\bi' \in \{0,1\}^K }
		\et^{K-\|\bi'\|_1}
		\no^{\|\bi'\|_1} \nonumber \\
	&\qquad \qquad  \prod_{\substack{k_2 \in [K]\\ i'_{k_2} = 0}}
		\big\| \D_{(k_2,0),\cS_{n_{k_2}}}	\big\|_2
		\prod_{\substack{k_2 \in [K]\\ i'_{k_2} = 1}}
		\big\| \D_{(k_2,1,l_{k_2}),\cS_{n_{k_2}}}	\big\|_2
		\bigg) \nonumber \\
	& +2 \bigg(
		\et^K
		\prod_{k_1 \in [K]}   \big\| \D_{(k_1,0),\cS_{n_{k_1}}}	\big\|_2
		\bigg) \bigg( \sum_{\substack{\bi' \in \{0,1\}^K \\ \|\bi'\|_1 \neq 0}}
		\et^{K-\|\bi'\|_1}
		\no^{\|\bi'\|_1}  \nonumber \\
	&\qquad \qquad \prod_{\substack{k_2 \in [K]\\ i'_{k_2} = 0}}
		\big\| \D_{(k_2,0),\cS_{n_{k_2}}}	\big\|_2
		\prod_{\substack{k_2 \in [K]\\ i'_{k_2} = 1}}
		\big\| \D_{(k_2,1,l_{k_2}),\cS_{n_{k_2}}}	\big\|_2
		\bigg) \nonumber \\
	&  \numrel{\leq}{r_e_2} 2\sqrt{s}  \bigg[
		\bigg( \sum_{k_1=0}^{K-1}
		{K \choose k_1}
		\et^{k_1}
		\no^{K-k_1}
		\bigg( \sqrt{\frac{3}{2}}\bigg)^{k_1}
		\bigg( \sqrt{\frac{1}{r^{2/K}}}\bigg)^{K-k_1}
		\bigg) \nonumber \\
	&\ \qquad \bigg( \sum_{k_2=0}^{K}
		{K \choose k_2}
		\bigg( \et\sqrt{\frac{3}{2}}\bigg)^{k_2}
		\bigg) +
		\bigg( \et \sqrt{\frac{3}{2}}\bigg)^K \nonumber \\
	& \  \qquad
		\bigg( \sum_{k_2=0}^{K-1}
		{K \choose k_2}
		\et^{k_2}
		\no^{(K-k_2)}
		\bigg( \sqrt{\frac{3}{2}}\bigg)^{k_2}
		\bigg( \sqrt{\frac{1}{r^{2/K}}}\bigg)^{K-k_2}
		\bigg) \bigg] \nonumber \\
	&\numrel{\leq}{r_no_et_1} 2\sqrt{\frac{s\eps'}{r^2}}
		\bigg( \sum_{k_1=0}^{K-1} {K \choose k_1}
		\bigg(\sqrt{\frac{3}{2}} \bigg)^{k_1} \bigg) \nonumber \\
	&\  \qquad
		\bigg( \bigg( \sum_{k_2=0}^{K} {K \choose k_2}
		\bigg( \sqrt{\frac{3}{2}}\bigg)^{k_2}
		\bigg) + \bigg( \sqrt{\frac{3}{2}}\bigg)^K
		\bigg) \nonumber \\
	& \numrel{\leq}{r_last} 3^{2K+1}\sqrt{\frac{s\eps'}{r^2}},
	\label{eq:sigma_diff_2}
	\end{align}
where \eqref{r_e_2} follows from \eqref{A2-B2_2}, \eqref{r_no_et_1} follows from replacing the value for $\no$ and the fact that $\et<1$, $\eps'/r^2<1$ (by assumption), and \eqref{r_last} follows from similar arguments in \eqref{eq:sigma_diff}. The rest of the proof follows the same arguments as in Lemma~\ref{lemma_Sp_I_UB} and \eqref{eq:MI_U_S} holds in this case as well.
\end{IEEEproof}

\begin{IEEEproof}[Proof of Theorem~\ref{thm:PrtCnvrs}]
Any dictionary $\D\in \cX(\I_p,r)$ can be written as
	\begin{align}
	\D& = \A \otimes  \B \nonumber\\
	&= (\I_{p_1} + \Delt_1) \otimes (\I_{p_2} + \Delt_2),
	\end{align}
We have to ensure that $\|\mb{D}-\mb{I}_p\|_F\leq r$. We have
	\begin{align}
	&\|\D-\I_p\|_F \nonumber \\
	&\qquad= \|\I_{p_1} \otimes \Delt_2
		+ \Delt_1\otimes \I_{p_2}
		+ \Delt_1\otimes \Delt_2\|_F  \nonumber\\
	&\qquad \leq \|\I_{p_1} \otimes \Delt_2\|_F
		+ \|\Delt_1\otimes \I_{p_2} \|_F
		+ \|\Delt_1\otimes \Delt_2\|_F  \nonumber\\
	&\qquad = \|\I_{p_1} \|_F\| \Delt_2\|_F
		+ \|\Delt_1\|_F\| \I_{p_2} \|_F
		+ \|\Delt_1\|_F\| \Delt_2\|_F  \nonumber\\
	&\qquad \leq r_2\sqrt{p_1} + r_1\sqrt{p_2} +r_1r_2 \nonumber \\
	&\qquad \numrel{\leq}{r_r1r2p}r,
	\end{align}
where \eqref{r_r1r2p} follows from \eqref{eq:PC_c_r}. Therefore, we have
	\begin{align}
	\D \in &\bigg\{\A \otimes  \B
	=(\I_{p_1} + \Delt_1) \otimes (\I_{p_2}
		+ \Delt_2)\big| \  \|\Delt_1\|	_F \leq r_1, \nonumber \\
	&\qquad \|\Delt_2\|_F \leq r_2,\ r_2\sqrt{p_1}
		+ r_1\sqrt{p_2} +r_1r_2 \leq r, \nonumber\\
	&\qquad \|\ba_{l_1}\|_2=1, l_1\in [p_1], \
		\|\bb_{l_2}\|_2=1, l_2\in [p_2]\bigg\}.
	\end{align}

In this case, the new observation vectors $\y'_{(n,j)}$ can be written as
	\begin{align} \label{eq:yp_xp}
	\y'_{(n,j)} = \A \x'_{(n,j)} + \A_p \x_n, \ j \in [p_2], \ n \in [N],
	\end{align}
where $\A_p \triangleq (\A\otimes \Delt_2)^{\mc{T}_n}$ denotes the matrix consisting of the rows of $(\A\otimes \Delt_2)$ with indices $	\mc{T}_n \triangleq ip_2+j$, where $i = \{0\} \cup [p_1-1]$ and $j = \big( (n-1) \bmod  p_2 \big) + 1$.

Similarly, for $\y''_{(n,j)}$ we have
	\begin{align} \label{eq:yz_xz}
	\y''_{(n,j)} = \B \x''_{(n,j)} + \B_p \x_n, \ j \in [p_1], \ n \in [N],
	\end{align}
where $\B_p \triangleq (\Delt_1 \otimes \B)^{\mc{I}_n}$ denotes the matrix consisting of the rows of $(\Delt_1 \otimes \B)$ with indices
$\mc{I}_n \triangleq jp_2+i$, where $ i = \{0\} \cup [p_2-1]$ and $j = (n-1) \bmod  p_1$.
Given the fact that $\x_n \in \{-1,0,1\}^p$, $\sigma_a^2=1$ and $\|\x_n\|_2^2=s$, after division of the coefficient vector according to \eqref{eq:xp} and \eqref{eq:xpp}, we have
	\begin{align} \label{eq:Ex}
	\bbE_{\x_n}\lr{x_{n,l}^2}
	&=\bbE_{\x_{(n,j_1)}'}\lr{ {x'}_{(n,j_1),l_1}^2 }
	=\bbE_{\x_{(n,j_2)}''}\lr{ {x''}_{(n,j_2),l_2}^2 } \nonumber \\
	&=\frac{s}{p},
	\end{align}
for any $n \in [N], j_1 \in [p_2], j_2 \in [p_1], l \in [p], l_1 \in [p_1]$, and $l_2 \in [p_2]$. The $\SNR$ is
	\begin{align} \label{eq:SNR_2}
	\SNR = \frac{\bbE_\x\lr{\|\x\|_2^2}}{\bbE_{\boldsymbol{\eta}}\lr{\|\boldsymbol{\eta}\|_2^2}} = \frac{s}{m\sigma^2}.
	\end{align}	
We are interested in upper bounding $\bbE_\Y\lr{\norm{\wh{\D}(\Y)-\D}_F^2 }$. For this purpose we first upper bound $ \bbE_\Y\lr{\norm{\wh{\A}(\Y)-\A}_F^2}$ and $ \bbE_\Y\lr{\norm{\wh{\B}(\Y)-\B}_F^2}$.
We can split these MSEs into the sum of column-wise MSEs:
	\begin{align}
	\bbE_\Y\lr{\norm{\wh{\A}(\Y)-\A}_F^2}
		= \sum_{l=1}^{p_1}\bbE_\Y\lr{\norm{\wh{\ba}_l(\Y)-\ba_l}_2^2}.
	\end{align}
By construction:
	\begin{align}
	\norm{\wh{\ba}_l(\Y)-\ba_l}_2^2
	&\leq 2\left(\norm{\wh{\ba}_l(\Y)}_2^2
		+ \norm{\ba_l}_2^2 \right)  \nonumber\\
	&\numrel{\leq}{r_xhat_n} 4 , \label{norm_u}
	\end{align}
where \eqref{r_xhat_n} follows from the projection step in \eqref{update_b}.
We define the event $\mc{C}$ to be
	\begin{align}
	\mc{C} \triangleq \bigcap_{\substack{n\in [N] \\ l\in [p]} } \lr{|\eta_{n,l}| \leq 0.4}.
	\end{align}
In order to find the setting under which $\bbP\lr{\wh{\X}=\X|\mc{C}}=1$,  i.e., when recovery of the coefficient vectors is successful, we observe the original observations and coefficient vectors satisfy:
	\begin{align}
	y_{n,l} - x_{n,l}
		=  \left(\I_{p_1} \otimes \Delt_2
			+ \Delt_1\otimes \I_{p_2}
			+ \Delt_1 \otimes \Delt_2 \right)^{l}\x_n
			+ \eta_{n,l}
	\end{align}
and
	\begin{align}
	&\left|\left(\I_{p_1} \otimes \Delt_2
		+ \Delt_1\otimes \I_{p_2}
		+ \Delt_1 \otimes \Delt_2 \right)^{l}\x_n
		+ \eta_{n,l} \right|
		\notag \\
	&\ \leq
		\left\|\left(\I_{p_1} \otimes \Delt_2
		+ \Delt_1\otimes \I_{p_2}
		+ \Delt_1 \otimes \Delt_2 \right)^{l}\right\|_2
		\left\|\x_n \right\|_2
		+ | \eta_{n,l} |  \nonumber\\
	&\ \leq
		\left( \| \Delt_1\|_F + \|\Delt_2\|_F + \| \Delt_1 \|_F \| \Delt_2 \|_F \right)\|\x_n\|_2
		+ | \eta_{n,l} |  \nonumber\\
	&\ \leq
		(r_1 + r_2 +r_1r_2)\sqrt{s} + | \eta_{n,l} |.
	\end{align}
By using the assumption $(r_1 + r_2 +r_1r_2)\sqrt{s}\leq 0.1$ and conditioned on the event $\mc{C}$, $|\eta_{n,l} |\leq 0.4$, we have that for every $n\in [N]$ and $l \in [p]$:
	\begin{align} \label{eq:cond_arg}
	\begin{cases}
    	y_{n,l} >0.5      & \quad \text{if } x_{n,l}=1,\\
    	-0.5<y_{n,l} <0.5 & \quad \text{if } x_{n,l}=0,\\
    	y_{n,l} <-0.5     & \quad \text{if } x_{n,l}=-1,
	\end{cases}
	\end{align}
thus, ensuring correct recovery of coefficients ($\wh{\X}=\X$) using the thresholding technique \eqref{t_alg} when conditioned on $\mc{C}$. Using standard tail bounds for Gaussian random variables~\cite[(92)]{jung2015minimax},~\cite[Proposition 7.5]{foucart2013mathematical} and taking a union bound over all $pN$ i.i.d. variables $\{\eta_{n,l}\}, n \in [N], l \in [p]$, we have
	\begin{align} \label{eq:P_Cc}
	\bbP\lr{\mc{C}^c} \leq \exp\left(-\frac{0.08pN}{\sigma^2}\right).
	\end{align}
	
To find an upper bound for $\bbE_\Y\lr{\|\wh{\ba}_l(\Y)-\ba_l\|_2^2}$, we can write it as
\begin{align}
	\bbE_\Y & \lr{\norm{\wh{\ba}_l(\Y)-\ba_l}_2^2}
		=	\bbE_{\Y,\N}\lr{\norm{\wh{\ba}_l(\Y)-\ba_l}_2^2|\mc{C}} \bbP(\mc{C}) \nonumber \\
	& +   \bbE_{\Y,\N}\lr{\norm{\wh{\ba}_l(\Y)-\ba_l}_2^2
		|\mc{C}^c} \bbP(\mc{C}^c)  \nonumber\\
	&\numrel{\leq}{r_C_e}  \bbE_{\Y,\N}
		\lr{\norm{\wh{\ba}_l(\Y)-\ba_l}_2^2|\mc{C}}
		+ 4 \exp\left(-\frac{0.08pN}{\sigma^2}\right),
	\end{align}
where \eqref{r_C_e} follows from \eqref{norm_u} and \eqref{eq:P_Cc}. To bound $\bbE_{\Y,\N}\lr{\norm{\wh{\ba}_l(\Y)-\ba_l}_2^2|\mc{C}}$, we have
	\begin{align}
	& \bbE_{\Y,\N} \lr{ \norm{\wh{\ba}_l(\Y)-\ba_l}_2^2|\mc{C} }
		\nonumber \\
	&\quad =	\bbE_{\Y,\N}\lr{\norm{P_{\mathcal{B}_1}(\wt{\ba}_l(\Y)	)
		-\ba_l}_2^2|\mc{C}} \nonumber\\
	&\quad \numrel{\leq}{r_proj_r} \bbE_{\Y,\N}
		\lr{\norm{\wt{\ba}_l(\Y)-\ba_l}_2^2|\mc{C}} \nonumber\\
	&\quad \numrel{=}{e_1} \bbE_{\Y,\N} \bigg\{ \bigg\| \frac{p_1}{Ns}
		\sum_{n=1}^{N} \sum_{j=1}^{p_2}\wh{x'}_{(n,j),l}\y_{(n,j)}'
		-\ba_l\bigg\|_2^2\bigg |\mc{C} \bigg\} \nonumber\\
	&\quad \numrel{=}{r_xhat_e_x} \bbE_{\Y,\X,\N}\bigg\{
		\bigg\|\frac{p_1}{Ns}
		\sum_{n=1}^{N} \sum_{j=1}^{p_2} x_{(n,j),l}'\y_{(n,j)}'
		-\ba_l\bigg\|_2^2\bigg|\mc{C} \bigg\} \nonumber\\
	&\quad \numrel{=}{ee_1} \bbE_{\X,\N}\bigg\{\bigg\|\frac{p_1}{Ns}
		\sum_{n=1}^{N} \sum_{j=1}^{p_2} x_{(n,j),l}' \big(\A\x_{(n,j)}'
		+\A_p \x_n \nonumber \\
	&\quad \qquad \qquad \qquad \qquad \qquad \qquad \qquad + \boldsymbol{\eta}_{(n,j)}'\big)  -\ba_l\bigg\|_2^2\bigg|\mc{C} \bigg\}
		\nonumber\\
	&\quad \numrel{\leq}{e_2} 2\bbE_{\X,\N}
		\bigg\{\bigg\|\frac{p_1}{Ns}
		\sum_{n=1}^{N} \sum_{j=1}^{p_2}
		 x_{(n,j),l}' \boldsymbol{\eta}_{(n,j)}'\bigg\|_2^2\bigg|\mc{C}\bigg\}
		 \nonumber \\
	&\ \quad + 4\bbE_{\X,\N}\bigg\{\bigg\|\ba_l-\frac{p_1}{Ns}
		\sum_{n=1}^{N} \sum_{j=1}^{p_2}  x_{(n,j),l}'
		\sum_{t=1}^{p_1} \ba_t x_{(n,j),t}'  \bigg\|_2^2\bigg|\mc{C}\bigg\} \nonumber \\
	&\ \quad +4\bbE_{\X,\N} \bigg\{\bigg\|\frac{p_1}{Ns}
		\sum_{n=1}^{N} \sum_{j=1}^{p_2}  x_{(n,j),l}'
		\sum_{t=1}^p \ba_{p,t} x_{n,t} \bigg\|_2^2 \bigg|\mc{C}\bigg\} ,		\label{eq:3_term}
	\end{align}
where \eqref{r_proj_r} follows from the fact that $\|\ba_l\|_2=1$, \eqref{e_1} follows from \eqref{eq:a_update}, \eqref{r_xhat_e_x} follows from the fact that conditioned on the event $\mc{C}$, $\widehat{\X} = \X$, \eqref{ee_1} follows from \eqref{eq:yp_xp} and \eqref{e_2} follows from the fact that $\|\x_1 +\x_2\|_2^2 \leq 2(\|\x_1\|_2^2 +\|\x_2\|_2^2)$.
We bound the three terms in \eqref{eq:3_term} separately. Defining $\nu \triangleq \mc{Q}(-0.4/\sigma) - \mc{Q}(0.4/\sigma)$, where $\mc{Q}(x)\triangleq \int_{z=x}^\infty \frac{1}{\sqrt{2\pi}}\exp(-\frac{z^2}{2})dz$, we can bound the noise variance conditioned on $\mc{C}$, $\sigma^2_{\eta_{n,t}}$, by~\cite{jung2015minimax}
   \begin{align} \label{eq:sig_nu}
   \sigma^2_{\eta_{n,t}} \leq \frac{\sigma^2}{\nu}.
   \end{align}

The first expectation in \eqref{eq:3_term} can be bounded by
	\begin{align} \label{E_1}
	&\bbE_{\X,\N} \lr{\bigg\|\frac{p_1}{Ns}
		\sum_{n=1}^{N} \sum_{j=1}^{p_2}
		 x_{(n,j),l}' \boldsymbol{\eta}_{(n,j)}'\bigg\|_2^2\bigg|\mc{C}}
		 \notag \\
	& = \left(\frac{p_1}{Ns}\right)^2
		\sum_{n,n'=1}^{N} \sum_{j,j'=1}^{p_2}
		\bbE_{\X,\N} \bigg\{{x'}_{(n,j),l}{x'}_{(n',j'),l} \nonumber \\
	&\qquad \qquad \qquad	\qquad \qquad \qquad \qquad
	\quad {\boldsymbol{\eta}'}_{(n',j')}^\top \boldsymbol{\eta}'_{(n,j)}|\mc{C}\bigg\} \nonumber \\
	& = \left(\frac{p_1}{Ns}\right)^2
		\sum_{n=1}^{N} \sum_{j=1}^{p_2}  \sum_{t=1}^{m_1}
		\bbE_{\X,\N} \lr{{x'}_{(n,j),l}^2|\mc{C}}
		\bbE_{\X,\N}\lr{{\eta'}^2_{(n,j),t}|\mc{C}} \nonumber \\
	&\numrel{=}{r_x_c} \left(\frac{p_1}{Ns}\right)^2 Np_2
	\bbE_{\X} \lr{{x'}_{(n,j),l}^2}
	\bbE_{\N}\lr{{\eta'}^2_{(n,j),t}|\mc{C}}  \nonumber\\
	&\numrel{\leq}{r_sig_x} \left(\frac{p_1}{Ns}\right)^2 Np_2\left(\frac{s}{p}\right) \left(\frac{m_1\sigma^2}{\nu} \right)\nonumber\\
	& \numrel{\leq}{r_nu_1} \frac{2 m_1 p_1\sigma^2}{Ns},
	\end{align}
where \eqref{r_x_c} follows from the fact that $\x_{(n,j)}'$ is independent of the event $\mc{C}$, \eqref{r_sig_x} follows from \eqref{eq:Ex} and \eqref{eq:sig_nu}, and \eqref{r_nu_1} follows from the fact that $\nu \geq 0.5$ under the assumption that $\sigma\leq 0.4$~\cite{jung2015minimax}.

To bound the second expectation in \eqref{eq:3_term}, we use similar arguments as in Jung et al.~\cite{jung2015minimax}. We can write
	\begin{align}
	& \bbE_\X\big\{x'_{(n,j),l}x'_{(n,j),t}x'_{(n',j'),l}x'_{(n',j'),t'}\big\} =
		\nonumber \\
	&\qquad \begin{cases}
    	(\frac{s}{p})^2 &\quad \text{if} \ (n,j)=(n',j')
    	\ \text{and} \ t=t'\neq l,\\                                                                                                                                                                                                                                                                                                                                                                                                                                                                                                                                                                                                                                                                                                                                                                                  		(\frac{s}{p})^2 &\quad \text{if} \ (n,j)\neq(n',j')
    	\ \text{and} \ t=t'=l,\\
    	\frac{s}{p}     &\quad \text{if} \ (n,j)=(n',j')   \ \text{and} \ t=t'=l,\\
    	0               &\quad \text{otherwise},
  		\end{cases} \label{eq:E_x4}
	\end{align}
and we have
	\begin{align} \label{E_2}
	& \bbE_{\X,\N}\bigg\{\bigg\|\ba_l-\frac{p_1}{Ns}
		\sum_{n=1}^{N} \sum_{j=1}^{p_2}  x_{(n,j),l}'
		\sum_{t=1}^{p_1} \ba_t x_{(n,j),t}'
		\bigg\|_2^2\bigg|\mc{C}\bigg\} \nonumber \\
	&\quad  \leq \ba_l^\top \ba_l - \frac{2p_1}{Ns}
		\sum_{n=1}^{N} \sum_{j=1}^{p_2} \sum_{t=1}^{p_1} \ba_l^\top \ba_t
		\bbE_\X\lr{x_{(n,j),l}'x_{(n,j),t}'} \nonumber \\
	&\quad \quad + \left(\frac{p_1}{Ns}\right)^2
		\sum_{n,n'=1}^{N}  \sum_{j,j'=1}^{p_2}
		\sum_{t,t'=1}^{p_1} \ba_{t'}^\top \ba_t
		\nonumber \\
	&\qquad \qquad \qquad \qquad \quad \bbE_\X\lr{x_{(n',j'),l}'x_{(n',j'),t'}'x_{(n,j),l}'x_{(n,j),t}'}
		\nonumber \\
	&\quad = 1 - \lrp{\frac{2p_1}{Ns}} \lrp{p_2N} \lrp{\frac{s}{p}}
		+ \left(\frac{p_1}{Ns}\right)^2\lrp{p_2N} \nonumber \\
	&\qquad \qquad \qquad
		\bigg(\frac{s}{p} +(p_1-1)\lrp{\frac{s}{p}}^2
		+(p_2N-1)\lrp{\frac{s}{p}}^2 \bigg) \nonumber\\
	&\quad = \frac{p_1}{N}\left( \frac{1}{s}
		+ \frac{1}{p_2}-\frac{2}{p}\right)   \nonumber \\
    &\quad \leq \frac{2p_1}{N}.
	\end{align}

To upper bound the third expectation in \eqref{eq:3_term}, we need to bound the $\ell_2$ norm of columns of $\A_p$. We have
	\begin{align} \label{eq:a_p_t_l2}
	\forall t \in [p] : \|\ba_{p,t}\|_2^2
	& \numrel{\leq}{r_ap_t} \|(\A\otimes \Delt_2)_t\|_2^2 \nonumber \\
	& \leq \|\ba_l\|_2^2 \|\Delt_2\|_F^2 \nonumber \\
	& = r_2^2,
	\end{align}
where $(\A\otimes \Delt_2)_t$ denotes the $t$-th column of $(\A\otimes \Delt_2)$ and \eqref{r_ap_t} follows from the fact that $\A_p$ is a submatrix of $(\A\otimes \Delt_2)$. Moreover, similar to the expectation in \eqref{eq:E_x4}, we have
	\begin{align}
	&\bbE_\X\big\{x'_{(n,j),l}x'_{(n',j'),l}x_{n,t}x_{n',t'}\big\} =
	\nonumber \\
	&\qquad \begin{cases}
    	(\frac{s}{p})^2 &\quad \text{if} \ (n,j)=(n',j')   \ \text{and} \ t=t'\neq l',\\                                                                                                                                                                                                                                                                                                                                                                                                                                                                                                                                                                                                                                                                                                                                                                                  		(\frac{s}{p})^2 &\quad \text{if} \ (n,j)\neq(n',j')\ \text{and} \ t=t'=l',\\
    	\frac{s}{p}     &\quad \text{if} \ (n,j)=(n',j')   \ \text{and} \ t=t'=l',\\
    	0               &\quad \text{Otherwise},
  		\end{cases}
	\end{align}
where $l'$ denotes the index of the element of $\x_n$ corresponding to $x_{(n,j),l}'$.
Then, the expectation can be bounded by
	\begin{align} \label{E_3}
	&\bbE_{\X,\N} \bigg\{\bigg\|\frac{p_1}{Ns}
		\sum_{n=1}^{N} \sum_{j=1}^{p_2}  x_{(n,j),l}'
		\sum_{t=1}^p \ba_{p,t} x_{n,t} \bigg\|_2^2 \bigg|\mc{C} \bigg\} \nonumber \\
	&\quad =  \lrp{\frac{p_1}{Ns}}^2
		\sum_{n,n'=1}^{N} \sum_{j,j'=1}^{p_2}
		\sum_{t,t'=1}^{p} 	\ba_{p,t'}^\top \ba_{p,t}
		\nonumber \\
	&\quad \qquad \qquad \qquad \qquad \qquad
		\bbE_{\X}\lr{ x_{(n,j),l}' x'_{(n',j'),l}x_{n,t}x_{n',t'} }\nonumber \\	
    &\quad \numrel{\leq}{r_a_p_l2} r_2^2 \lrp{\frac{p_1}{Ns}}^2
    	Np_2 \bigg( \frac{s}{p} + (p-1)\lrp{\frac{s}{p}}^2 \nonumber \\
    &\qquad \qquad \qquad \qquad \qquad \qquad \qquad \quad
    	 + (Np_2-1)\lrp{\frac{s}{p}}^2 \bigg)   \nonumber \\
    &\quad \leq r_2^2 \lrp{ \frac{p_1}{Ns} + \frac{p_1}{N}+1 } \nonumber \\
    &\quad \numrel{\leq}{r_r_2_p} \frac{p_1}{N},
	\end{align}
where \eqref{r_a_p_l2} follows from \eqref{eq:a_p_t_l2} and \eqref{r_r_2_p} follows from the assumption in \eqref{eq:PC_c_r}. Summing up \eqref{E_1}, \eqref{E_2}, and \eqref{E_3}, we have
	\begin{align}
	&\bbE_\Y\lr{\norm{\wh{\ba}_l(\Y)-\ba_l}_2^2} \nonumber \\
	&\qquad \quad \leq \frac{4p_1}{N}\left( \frac{m_1\sigma^2}{s} + 3\right)+
		4\exp\left(-\frac{0.08pN}{\sigma^2}\right).
	\end{align}
Summing up the MSE for all columns, we obtain:
	\begin{align} \label{A_error}
	&\bbE_\Y\lr{\norm{\wh{\A}(\Y)-\A}_F^2} \nonumber \\
	&\quad \quad  \leq	\frac{4p_1^2}{N}\left( \frac{m_1\sigma^2}{s} + 3\right)
		+4p_1\exp\left(-\frac{0.08pN}{\sigma^2}\right).
	\end{align}
We can follow similar steps to get
	\begin{align} \label{B_error}
	&\bbE_\Y\lr{\norm{\wh{\B}(\Y)-\B}_F^2} \nonumber \\
	&\quad \quad \leq
		\frac{4p_2^2}{N}\left( \frac{m_2\sigma^2}{s} + 3\right)
		+4p_2\exp\left(-\frac{0.08pN}{\sigma^2}\right).
	\end{align}
From \eqref{A_error} and \eqref{B_error}, we get
	\begin{align}
	&\bbE_\Y \lr{\norm{\wh{\D}(\Y)-\D}_F^2  }
	\notag \\
	&\quad = \bbE_\Y\lr{\norm{\wh{\A}(\Y)\otimes \wh{\B}(\Y) -\A \otimes \B}_F^2}  \nonumber\\
	&\quad = \bbE_\Y \lr{\norm{(\wh{\A}(\Y) -\A)\otimes \wh{\B}(\Y)
		+ \A \otimes (\wh{\B}(\Y)-\B)}_F^2}  \nonumber\\
	&\quad \leq 2\bigg(\bbE_\Y \lr{\norm{(\wh{\A}(\Y) -\A)\otimes \wh{\B}(\Y) }_F^2 } \nonumber \\
	&\quad \qquad + \bbE_\Y\lr{ \norm{\A \otimes (\wh{\B}(\Y)-\B) }_F^2  }\bigg)  \nonumber\\
	&\quad \leq 2\bigg( \bbE_\Y \lr{ \norm{(\wh{\A}(\Y) -\A)}_F^2} \bbE_\Y
		\lr{ \norm{ \wh{\B}(\Y) }_F^2} \nonumber \\
	&\quad \qquad + \norm{\A }_F^2 \bbE_\Y \lr{ \norm{ (\wh{\B}(\Y)-\B)}_F^2 } \bigg)  \nonumber\\
	&\quad \leq 2\bigg(p_2 \bbE_\Y \lr{ \norm{(\wh{\A}(\Y) -\A)}_F^2}
		\nonumber \\
	&\quad \qquad +p_1\bbE_\Y \lr{ \norm{(\wh{\B}(\Y)-\B)}_F^2} \bigg)
		\nonumber\\
	&\quad \leq \frac{8p}{N}\bigg( \frac{\sigma^2}{s}\sum_{k=1}^2m_kp_k
		+3\sum_{k=1}^2 p_k\bigg)
		+ 8p\exp\left( -\frac{0.08pN}{\sigma^2} \right) \nonumber\\
	&\quad \numrel{=}{r_SNR_2}\frac{8p}{N}\bigg( \frac{\sum_{k=1}^2m_kp_k}{m\SNR}+3\sum_{k=1}^2 p_k \bigg)
		+ 8p\exp\left( -\frac{0.08pN}{\sigma^2} \right),
	\end{align}
where \eqref{r_SNR_2} follows from \eqref{eq:SNR_2}.
\end{IEEEproof}

%
%

\bibliographystyle{IEEEtran}
\bibliography{IEEEabrv,refs_full}
\newpage

\begin{IEEEbiographynophoto}
{Zahra Shakeri} is pursuing a Ph.D. degree at Rutgers University, NJ, USA. She is a member of the INSPIRE laboratory. She received her M.Sc. degree in Electrical and Computer Engineering from Rutgers University, NJ, USA, in 2016 and her B.Sc. degree in Electrical Engineering from Sharif University of Technology, Tehran, Iran, in 2013.
Her research interests are in the areas of machine learning, statistical signal processing, and multidimensional data processing.
\end{IEEEbiographynophoto}

\begin{IEEEbiographynophoto}
{Waheed U. Bajwa} received BE (with Honors) degree in electrical engineering from the National University of Sciences and Technology, Pakistan in 2001, and MS and PhD degrees in electrical engineering from the University of Wisconsin-Madison in 2005 and 2009, respectively. He was a Postdoctoral Research Associate in the Program in Applied and Computational Mathematics at Princeton University from 2009 to 2010, and a Research Scientist in the Department of Electrical and Computer Engineering at Duke University from 2010 to 2011. He is currently an Associate Professor in the Department of Electrical and Computer Engineering at Rutgers University. His research interests include statistical signal processing, high-dimensional statistics, machine learning, networked systems, and inverse problems.

Dr. Bajwa has received a number of awards in his career including the Best in Academics Gold Medal and President’s Gold Medal in Electrical Engineering from the National University of Sciences and Technology (2001), the Morgridge Distinguished Graduate Fellowship from the University of Wisconsin-Madison (2003), the Army Research Office Young Investigator Award (2014), the National Science Foundation CAREER Award (2015), Rutgers University's Presidential Merit Award (2016), Rutgers Engineering Governing Council ECE Professor of the Year Award (2016, 2017), and Rutgers University's Presidential Fellowship for Teaching Excellence (2017). He is a co-investigator on the work that received the Cancer Institute of New Jersey's Gallo Award for Scientific Excellence in 2017, a co-author on papers that received Best Student Paper Awards at IEEE IVMSP 2016 and IEEE CAMSAP 2017 workshops, and a Member of the Class of 2015 National Academy of Engineering Frontiers of Engineering Education Symposium. He served as an Associate Editor of the IEEE Signal Processing Letters (2014 – 2017), co-guest edited a special issue of Elsevier Physical Communication Journal on ``Compressive Sensing in Communications" (2012), co-chaired CPSWeek 2013 Workshop on Signal Processing Advances in Sensor Networks and IEEE GlobalSIP 2013 Symposium on New Sensing and Statistical Inference Methods, and served as the Publicity and Publications Chair of IEEE CAMSAP 2015 and General Chair of the 2017 DIMACS Workshop on Distributed Optimization, Information Processing, and Learning. He is currently Technical Co-Chair of the IEEE SPAWC 2018 Workshop and serves on the MLSP, SAM, and SPCOM Technical Committees of the IEEE Signal Processing Society.
\end{IEEEbiographynophoto}

\begin{IEEEbiographynophoto}
{Anand D. Sarwate}(S'99--M'09--SM'14) received the B.S. degrees in electrical engineering and computer science and mathematics from the Massachusetts Institute of Technology, Cambridge, MA, USA, in 2002, and the M.S. and Ph.D. degrees in electrical engineering from the Department of Electrical Engineering and Computer Sciences (EECS), University of California, Berkeley (U.C. Berkeley), Berkeley, CA, USA.

He is a currently an Assistant Professor with the Department of Electrical and Computer Engineering, The State University of New Jersey, New Brunswick, NJ, USA, since January 2014. He was previously a Research Assistant Professor from 2011 to 2013 with the Toyota Technological Institute at Chicago; prior to this, he was a Postdoctoral Researcher from 2008 to 2011 with the University of California, San Diego, CA.

His research interests include information theory, machine learning, signal processing, optimization, and privacy and security. Dr. Sarwate received the NSF CAREER award in 2015, and the Samuel Silver Memorial Scholarship Award and the Demetri Angelakos Memorial Award from the EECS Department at U.C. Berkeley. He was awarded the National Defense Science and Engineering Graduate Fellowship from 2002 to 2005. He is a member of Phi Beta Kappa and Eta Kappa Nu.
\end{IEEEbiographynophoto}
\end{document}